# Fabrication and calibration of search coils

*M. Buzio*
CERN, Geneva, Switzerland

**Abstract**
In this paper the techniques available to make and calibrate magnetic search coils are reviewed, with emphasis on harmonic coil systems as the commonly-used optimal choice for integral measurements of accelerator magnets in terms of measuring range, accuracy, and cost. The topics treated, drawing extensively on half a century of experience at CERN, include mechanical and electrical design criteria, practical fabrication techniques, metrological considerations, and various calibration methods for coil parameters such as surface area, rotation radius, tilt angle etc. in both static or time-varying magnetic fields.

## 1 Introduction

### 1.1 Scope

This lecture is concerned with the design, construction, and calibration of search coils for the measurement of accelerator magnets. This well-established type of sensor represents in many practical cases the best choice in terms of range of application, accuracy and cost-effectiveness, especially where the field map over large volumes is sought. Sensors typically used for point-like measurements, such as Hall-effect plates or NMR probes are described in other lectures of this school (see Refs. [1] and [2] respectively).

The material presented here is drawn mostly from the author's experience within CERN's magnetic measurement team, which has been accumulating, over more than five decades, the instruments and know-how needed to characterize an immense variety of magnets for the many accelerators of the complex. By sharing this experience, this lecture aims at outlining the most important techniques that have to be mastered by anyone who wants to build or simply use correctly a search coil, with special emphasis on the practical aspects of the work. As the variety of requirements and parameters is wide, coils of the type described here can hardly be found off-the-shelf, and it is hoped that the topics discussed will help interested readers to choose and implement the technique that works best for them. More information can be found in earlier CERN Accelerator Schools almost exclusively devoted to the subject of magnetic measurements [3, 4].

### 1.2 Working principle

*Search coil* is the generic name of a widespread class of sensors consisting of one or more loops of conducting wire, exposed to a magnetic field **B** and generating an output voltage $V_C$ via Faraday's induction law (see Ref. [5] for a comprehensive general review). Let us consider a coil made with $N_T$ closed turns spanning an area $A$ with normal unit vector **n** and boundary $\partial A$ [1], as shown in Fig. 1. The output voltage is given by the total rate of change of the linked magnetic flux $\Phi$ as in Eq. (1), where the negative sign (a.k.a. Lenz's law) means that the induced e.m.f. tends to generate a current which, by the right-hand rule, gives rise to a field opposing the variation of the flux.

---

[1] Practical coils are usually planar; however, the shape can in principle be arbitrary. Also, note that the loop will always be closed as the voltage terminals need to be connected to a voltmeter.

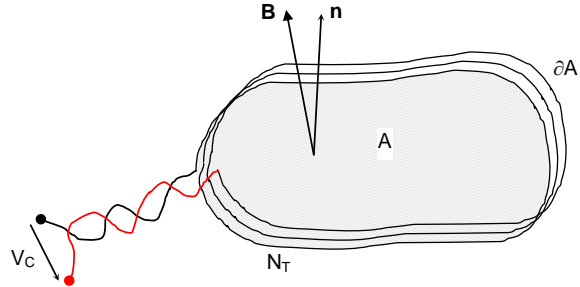

**Fig. 1:** Schematic representation of a search coil

$$V_C = -\frac{d\Phi}{dt} = -\frac{d}{dt}\iint_A \mathbf{B} \cdot \mathbf{n} \, dA = -\iint_A \frac{\partial \mathbf{B}}{\partial t} \cdot \mathbf{n} \, dA - \oint_{\partial A} \mathbf{v} \times \mathbf{B} \, d\ell \; . \qquad (1)$$

The r.h.s. of Eq. (1) implies that an output voltage can be generated in two ways:

- by a time-varying field, typically measured with a fixed coil (known in this case as a *flux loop*);
- by moving and/or deforming the coil with local velocity $\mathbf{v}$. This is usually achieved by rigid translation (in non-uniform fields only), rigid rotation (in which case we speak more properly of *harmonic coil*), or by forming a variable-geometry loop using for instance a stretched wire suspended between two translation stages [6].

If the geometry and position of the coil are known, the average value of the field over the coil's area can be inferred from either the instantaneous or the time-integrated voltage:

$$\Phi - \Phi_0 = -\int_0^t V_C \, dt \; . \qquad (2)$$

The advantage of using integration lies in the fact that the measured volt-seconds correspond directly to magnetic flux variations between the start and end configurations, irrespective of the irregularities of the path followed; moreover, the integration automatically filters out unwanted high-frequency noise components. On the other hand, any systematic error at the integrator's input, such as a voltage offset, gives rise to a drift which increases inexorably with time and may affect the accuracy of the result.

## 1.3 Requirements

In general, magnetic measurements are an essential step at various stages in the lifetime of accelerator magnets. In the prototyping phase, measurement results are used to verify design calculations, material properties, and fabrication methods; during series production the main aim is to monitor the quality of the manufacturing process; finally, throughout the operation of the machine, spare, reference, or refurbished magnets often have to be re-measured to assess their response to new operating conditions (current cycling, environmental conditions) or time-dependent effects such as those due to eddy currents or ageing.

In the common case where beam optics can be approached via the thin-lens approximation (magnet length << betatron oscillation wavelength [7]) the main target of a magnetic measurement is the integral field quality. This may include the strength of the main component, harmonic content (or, equivalently, uniformity of the main component), field direction, and location of the magnetic centre. All these quantities, in the most general case, have to be provided as a function of current excitation and history, ramp rate, and time. Local field quality, restricted by definition to a small fraction of the magnet's length, may also be of interest for several reasons, e.g., for spotting manufacturing or material defects [8], or for mapping the field as needed, for instance, in the case of spectrometers. High-resolution dynamic field measurements are also often used indirectly to localize and follow the propagation of quenches in superconducting magnets, as the redistribution of current in the magnet coil's cross-section causes detectable field harmonic changes [9].

These measurements have typically to conform to exacting specifications: long and narrow magnet gaps; main fields ranging from ~$10^{-4}$ T (typical residual field level) to about ~10 T (in superconducting magnets); relative accuracy of the order of $10^{-4}$ for the main field and $10^{-6}$ for the field errors. Given these constraints, slender search coils are often the natural and most cost-effective choice, especially so for integral measurements where harmonic (rotating) coils provide directly a field description in the format required by beam optics.

## 2    Circuit model

### 2.1    General case

In general, a search coil can be represented as the secondary winding of a transformer circuit where the primary is the magnet generating the flux, as the circuit scheme in Fig. 2 shows. The coil is therefore characterized by the following parameters:

- $L_c$ = self-inductance, depending on the coil geometry only
- $L^*$ = mutual inductance, depending on the coil geometry and the distribution of field (hence on position and orientation of the coil with respect to the magnet, field level as the magnet approaches saturation, ramp rate in case of eddy currents that modify the field pattern, etc.)
- $R_C$ = electrical resistance, depending on the coil geometry and material (and, to a small extent, on field level via magneto-resistance)
- $C_C$ = self-capacitance, depending on the geometry and the dielectric properties of the insulation
- $C^*$ = parasitic capacitance, depending on the geometry and the relative position w.r.t. nearby conductors such as the magnet coil and yoke, external structures, ground planes, etc.

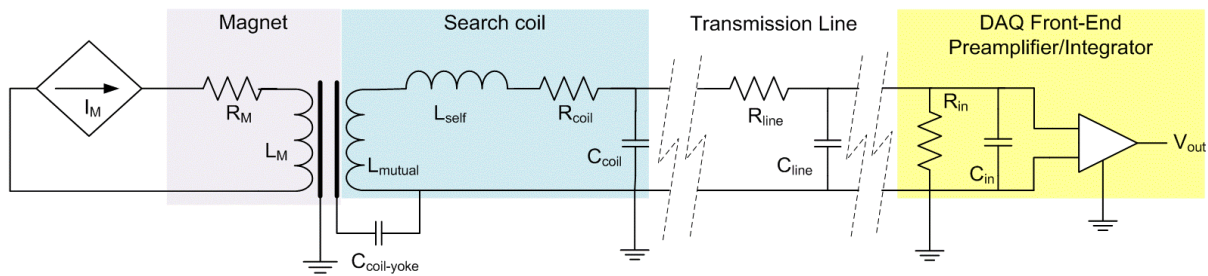

**Fig. 2**: Circuit diagram of a typical search coil measurement setup.
The magnet is supplied by the current source $I_M$.

The output voltage of the coil is fed to a data acquisition system (DAQ) via a transmission line, typically a twisted-pair or a coaxial cable. The acquisition, usually a digital integrator or an ADC/voltmeter, is mainly characterized by its own input impedance which should be as high as possible to minimize the current circulating in the coil.

### 2.2    Low-frequency approximation

A complete circuit description like the one seen above is fortunately unnecessary in the majority of practical cases as long as the spectral content of the signal remains sufficiently below the resonant frequencies (this condition is typically satisfied below 10–100 kHz, which excludes for example very fast pulsed linac or kicker magnets). In this case, one may use a low-frequency approximation (see Fig. 3) where the magnet coupling is replaced by a simple voltage source and the only meaningful parameters are the resistance of the coil and of the input stage of the DAQ. The resulting output tension $V_{in}$ and parasitic current $I_{coil}$ are given by

$$V_{in} = \underbrace{\frac{1}{1+\frac{R_{coil}}{R_{in}}}}_{k_R} V_{coil}, \qquad I_{coil} = \frac{V_{coil}}{R_{coil}+R_{in}}. \qquad (3)$$

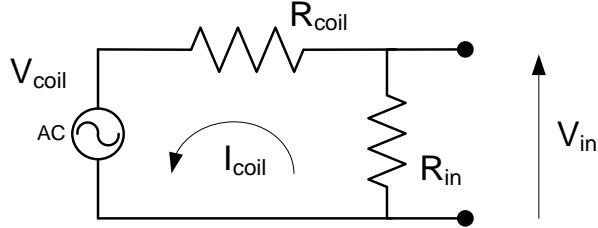

**Fig. 3:** Simplified search coil circuit in the low-frequency approximation. The inductive coupling to the magnet is represented by an AC voltage source and transmission line effects are neglected.

For typical input resistances in the $10^6$ Ω range and coil resistances of the order of a few ohm, parasitic currents $I_C$ of the order of a few $10^{-6}$ Ω are expected. These may have the following detrimental effects on the accuracy of the measurements:

- the measured voltage is smaller than the real coil voltage by the factor $k_R = 1+R_c/R_{in}$, which must be taken into account if high accuracy is desired;
- especially at high frequencies, the measurement might be perturbed by additional voltages due to the coil self-inductance and capacitance.

The factor $k_R$ can be used to apply a correction to the results, however, care must be taken to consider the possible external influences on the values of resistance such as operating temperature (for copper, $\partial R/R\partial T \approx 0.004/°C$) or magneto-resistive effects. The major factor affecting $R_{in}$ is normally the gain of the internal preamplifier stage, although one has also to expect a degree of dependency upon the frequency content of the signal. Joule heating effects in the coil due to $I_C$, on the other hand, are normally completely negligible (for example, an air-cooled 30 μm wire can carry indefinitely 5 mA without appreciable temperature rise).

## 2.3 Dipole-compensating coils

As explained in detail in another lecture of this School [10], accurate higher harmonic measurements require the use of an array of coils connected in such a fashion as to suppress the signal of the main field component. This 'bucking' or 'compensated' configuration not only improves the S/N ratio of the field error signal; crucially, it also abates the sensitivity to mechanical and geometrical imperfections of the rotating system. As an example, Fig. 4 shows a dipole compensation scheme in which two equal coils are mounted on the same rotating shaft parallel to each other. Two acquisition channels are run in parallel: coil 1 provides the so-called absolute signal, containing essentially the main dipole harmonic, while coil 2 is connected in series opposition to obtain the compensated signal which, ideally, contains only contributions from the error field. Assuming equal parameters for the two channels, absolute and compensated voltages can be calculated from:

$$V_{in}^{ABS} \cong \frac{1}{1+2\frac{R_C}{R_{in}}}V_{C1} + \frac{R_C}{R_{in}}V_{C2}, \qquad V_{in}^{CMP} \cong \frac{1}{1+3\frac{R_C}{R_{in}}}(V_{C1}-V_{C2}) - \frac{R_C}{R_{in}}V. \qquad (4)$$

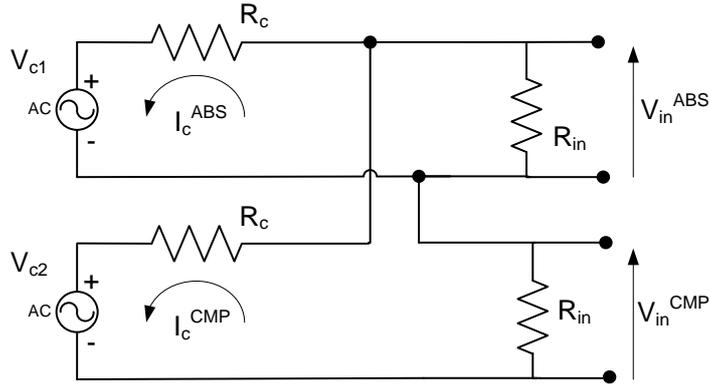

**Fig. 4:** Circuit diagram of two identical coils in a dipole-compensating configuration

Assuming for simplicity that $R_c/R_{in} \ll 1$ and that higher harmonics are small compared to the dipole (so that $V_{c1} \approx V_{c2}$) we find that also in this case the measured absolute signal is reduced by a factor $(1+R_c/R_{in})$. The compensated signal, ideally given by the difference $(V_{c1}-V_{c2})$, contains in fact an additional contribution from $V_{c2}$ that practically reintroduces the suppressed main harmonic even in case of geometrically perfect compensation. Analogous compensation schemes with two, three, or four coils are commonly used for quadrupole measurements, where the objective is to cancel out the dipole and quadrupole components using an appropriate combination of coil areas and rotation radii. Compensation of higher-order harmonics is in theory possible but seldom implemented, owing to the extreme mechanical accuracies needed. In all cases, the quality of the compensation can often be improved by adding appropriately dimensioned resistors in series to the coils, and modifying Eq. (4) accordingly.

## 3 Coil design criteria

### 3.1 Main technological options

In the majority of cases a search coil for accelerator magnets has a rectangular, elongated shape which, among other things, helps to achieve a precise geometry and simplifies the calculation of calibration coefficients. The circular shape, by contrast, is far easier to make and is very common in the wider context of inductors for electronics. Search coils of widely ranging dimensions are referenced in the literature, from sub-millimetre sizes for point-like measurements to several metres for geomagnetic and RF applications. The length range of interest for our application normally lies from a few centimetres to one metre or so, longer magnets often being measured by a sequence of smaller coils.

The main technologies available for winding cores are listed in Table 1 along with their main advantages and drawbacks. In the size range of interest the approach we have found most convenient consists in winding a mono- or multi-filament wire around a thin rigid rectangular form, which is in turn fastened to a suitable support, usually a cylindrical shaft, permitting rotation or translation. We predominantly use individual G10 forms which can be produced relatively cheaply in large quantities (see Fig. 5). As an alternative, the winding core can be obtained directly on the rotating shaft by carving out a series of parallel grooves, as shown for example in Ref. [11]. In this case the mechanical accuracy and stability of the measurement head is improved, at the cost of greater difficulty in the process of winding or repairing the coils. This technology may be the only practical choice, for instance, in the case of small rotating shafts, where there is no room for a traditional type of assembly.

**Table 1:** Summary of main coil fabrication techniques

| Coil Type | Pros | Cons |
| --- | --- | --- |
| Separate winding form | relatively easy to machine, handle, and wind accurately<br>large series production feasible<br>allows sorting for bucking<br>easy replacement if needed | need rigid support<br>long-term stability problems |
| Monolithic support | accurate and stable | winding many turns more difficult<br>potentially costlier |
| Printed Circuit Board (PCB) coils | extremely precise geometry<br>high compensation order possible<br>cheap in large runs | size limitations<br>calibration coefficients need geometrical corrections (end effects, large aspect ratio conductor) |
| Air-core | cheap and easy to make | small sizes only<br>poor geometrical accuracy |

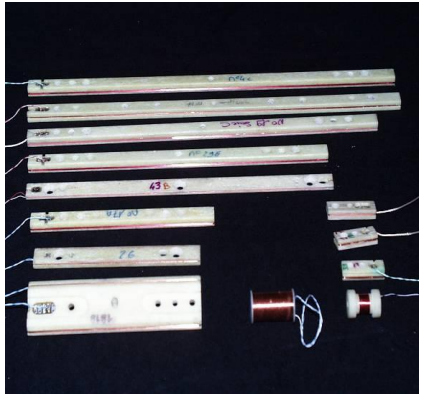

**Fig. 5:** A sample of CERN-made search coils of various shapes and sizes

### 3.2 Conducting wire

In this section we shall discuss the general criteria guiding the choice of a particular kind of wire, considering especially standard coils made by winding a single-filament wire in a spiral pattern. In the following sections we shall discuss the two major available alternatives, i.e., multi-filament and Litz wires.

Search coils are commonly wound from standard insulated copper wire, widely available on the market in gauges down to few hundredths of a millimetre. The next best metal in terms of conductivity, ductility and cost, i.e., aluminium, becomes competitive only for those current-carrying coils where the conductivity-to-weight ratio is important (aluminium's is about twice that of copper's).

The first real choice to be made concerns the shape of the wire cross-section, which could be either round or square. While a square shape could in principle allow more densely packed turns, it has two serious drawbacks: first, gauges smaller than 0.1~0.2 mm are difficult to find; second, and more important, with a square shape the unavoidable twist introduced during winding may result in a highly irregular geometry, detrimental to the accuracy of the measurement [12]. For these reasons, in our own work a round wire section has always been preferred.

The next major design choice is the diameter of the wire, which is the result of a compromise between two contrasting requirements: 1) a small cross-section to pack turns densely, thus saving space and approximating the geometrical ideal of a point-like cross section; 2) a large cross-section to increase mechanical strength and decrease electrical resistance (see Section 3.4). During winding, the wire has to be pulled taut to better adhere to the support; the ideal tension is just below yield, at which point the wire would neck and thus spoil the geometrical quality of the coil. Commercial wires are available in sizes normalized according to different standards, some of the most popular being the American Wire Gauge (AWG)[13] and the IEC 60228 [14].

The effective wire size must also include the necessary insulation, taking into account that the dielectric strength is approx. 20 V/µm for PVC and 8 V/µm for polyurethane. While in normal operating conditions coil voltages are by design of the order of a only few volts, off-normal events such as superconducting magnet quenches or power supply trips may give rise to dB/dt rates up to 10–100 T/s and hence coil voltages up to several hundred volts. To give an idea, common polyurethane-coated wire has an insulation thickness between 3.5 µm and 9 µm according to standard IEC 60317 [15], corresponding to the capacity to withstand a (rather conservative) inter-turn voltage drop of at least 30 V.

**3.3   Materials**

The material used to make the winding core and the coil support must have the following characteristics:

- high stiffness: to minimize measurement errors due to deformation and vibrations
- low thermal expansion coefficient
- mechanically stable, with a hard, non-porous surface to minimize moisture absorption
- non-magnetic: to avoid a) perturbations of the field to be measured and b) magnetic forces
- non-conducting: to avoid the occurrence of eddy currents and the consequent field perturbations and magnetic forces (NB: all measurement methods, whether the coil is moving or stationary, imply a flux change and therefore an e.m.f.)

The materials most commonly used are listed in Table 2 along with their main properties. It should be noted that good mechanical properties are generally in conflict with machinability, with the notable exception of Macor. This is a boro-silicate glass ceramic, loaded with mica to inhibit crack propagation, which associates high machinability with standard tools (achievable tolerances down to one µm) to high stiffness and low thermal expansion. Its main disadvantage is the high price, plus the difficulty of finding monolithic pieces longer than a few hundred millimetres on the market.

In most practical cases, the default choice is glass-reinforced epoxy laminates which are cheap, reasonably rigid and stable and allow tolerances down to ~0.01 mm, provided coated tools are used and frequently replaced. Among the standard grades classified by the National Electrical Manufacturer's Association (NEMA), G10 is the one with the highest hardness and the lowest thermal expansion. It is sensible to discuss the appropriate quality grade with the supplier, as the type of fibres, the chemical characteristics of the impregnation, etc. can make a great difference to the finished result. One must take into account the possible effects of the anisotropy of the composite, which, for example, should never be machined in the direction parallel to the fibres, lest the subsequent relaxation of in-built stresses warp irremediably the part. As another example, tubes fabricated with the fibres helically wound in one direction only can have nasty side-effects, such as unexpected torsion induced by temperature changes.

**Table 2:** Physical properties of most common materials for coil forms and supports

|  | Density $\rho$ [kg/m$^3$] | Young's modulus $E$ [GPa] | Thermal expansion $\alpha$ @ 300 K [ppm/K] | Resistivity[a] $\sigma$ [$\Omega$m] | Dielectric constant $\varepsilon_r$ [-] | Magnetic[b] susceptibility $\chi_m$ [-] |
|---|---|---|---|---|---|---|
| **Macor™** | 2520 | 64 | 0.9 | >10$^{14}$ | 6 | <10$^{-5}$ |
| **Vycor™ (96% Si)** | 2180 | 66 | 0.8 | >10$^{14}$ | 3.8 | <10$^{-5}$ |
| **Quartz (fused Si)** | 2200 | 72 | 0.6 | >10$^{14}$ | 3.8 | <2·10$^{-7}$ |
| **Carbon fibre** | 1600 | 250 | 6.5 | 10$^{-5}$ | n.a. | −1.6·10$^{-5}$ |
| **Ultem™ 2300** | 1510 | 5.5 | 20 | 10$^{17}$ | 30 | <10$^{-5}$ |
| **G10** | 1820 | 25 | 10 | >10$^{14}$ | 5.2 | <10$^{-5}$ |
| **Al$_2$O$_3$** | 3980 | 380 | 6.5 | >10$^{14}$ | 9.1 | <10$^{-5}$ |

[a] All materials are practically perfect insulators, except for carbon fibre. [b] All are also non-magnetic.

At the two extremes of the performance scale we find glass-reinforced plastics like Ultem, cheap and easy to machine but rather poor mechanically, and sintered ceramics such as alumina (Al$_2$O$_3$), with outstanding rigidity and stability but exceedingly difficult to handle, hence expensive. While raw alumina powder is very cheap, being the basis for a number of large-volume industrial applications (production of aluminium, refractory materials, abrasives, electrical insulators etc.), the sintered material is very hard and fragile and so it has to be ground with special techniques using diamond-coated tools (tolerances can be a few micrometres). Only a few firms in Europe are able to produce large parts; for instance, the 1.3 m long tubes used for the LHC rotating coil system (see Section 5) had to be built in three parts and then glued together. Other materials are normally reserved for very specific cases: quartz is also difficult to machine, but useful for cryogenic measurements due to its thermal stability; or carbon fibre which is slightly conductive, but has the highest stiffness-to-weight ratio.

### 3.4 Design of rectangular coils

We shall now consider the design criteria for a rectangular-shaped coil, by far the most common case. We will assume an ideal geometry as shown in Fig. 6, where the transverse size of the windings is negligible w.r.t. the size of the coil.

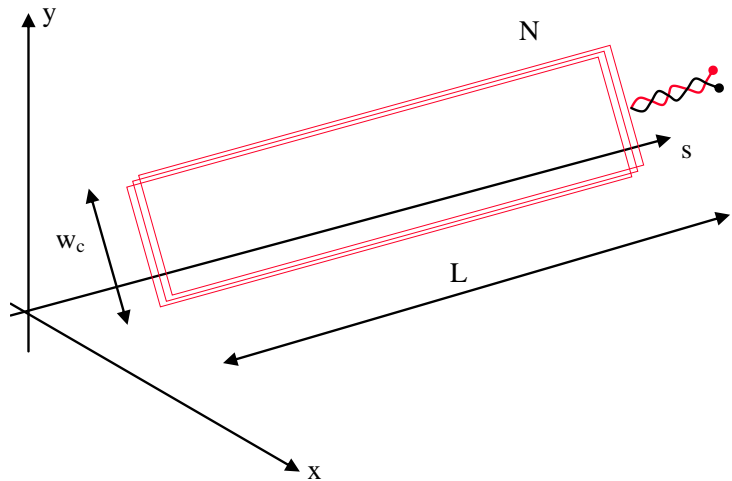

**Fig. 6:** Ideal geometry of a rectangular search coil

The main design parameters are listed below:

1) $R_C$ = coil resistance. As seen above, a low resistance minimizes the error due to a finite input stage impedance as well as the thermal noise $\sqrt{4k_b RT\Delta f}$.
2) $L_C$ = coil self-impedance. This parameter is important only in the high-frequency regime.
3) $L$ = coil length. This parameter is normally dictated by the size of the magnet being measured and by the nature of the information sought, e.g., integral or local field.
4) $w_C$ = coil width. This parameter too is often fixed by geometrical constraints (e.g., a coil of width $w_C$ rotating at a radius $R \gg w_C$ will be blind to the harmonic of order $2\pi R/w_C$).
5) $N_T$ = number of turns.
6) $\varnothing_W$ = wire diameter.
7) $A_C$ = coil area exposed to flux change. The area should be dimensioned to obtain an output voltage as large as possible, although not exceeding the input range of the acquisition electronics (typically, 5 or 10 V peak-to-peak) to avoid saturation of the input stage (this can have nasty side-effects such as a slow recovery of the nominal performance of the electronics). The output voltage can be derived from Eq. (1) for three canonical cases as follows:

$$-V_c = \frac{\partial \Phi}{\partial t} = \begin{cases} A_c \dot{B} & \text{Fixed coil in a time-varying field} \\ A_c B\omega & \text{Coil rotating at angular speed } \omega \text{ in a constant field} \\ A_c \nabla B v & \text{Coil translating at speed } v \text{ in a constant field with a gradient} \end{cases} \quad (5)$$

Note that the calculation of peak voltage must include a reasonable safety factor, at least 2~3, to takes into account the unavoidable voltage fluctuations due to irregularities of motion, power supply ripple, mechanical or electrical noise etc.

These seven parameters are linked by three equations:

$$\begin{cases} A_c = N_T \ell_c w_c \\ R_c = \frac{8}{\pi} N_T \rho \frac{\ell_c + w_c}{\varnothing_w^2} \\ L_c = \frac{\mu_0}{\pi} N_T^2 \left( \ell_c \ln \frac{\ell_c}{\varnothing_w} + w_c \ln \frac{w_c}{\varnothing_w} + 2\sqrt{\ell_c^2 + w_c^2} - \ell_c \sinh^{-1} \frac{\ell_c}{w_c} - w_c \sinh^{-1} \frac{w_c}{\ell_c} - \frac{7}{4}(\ell_c + w_c) \right) \end{cases} \quad (6)$$

In other words, one may fix four parameters and the other three will follow. As an example of parametrization, Fig. 7 shows what one obtains by fixing coil width and wire diameter, and considering area and length as independent.

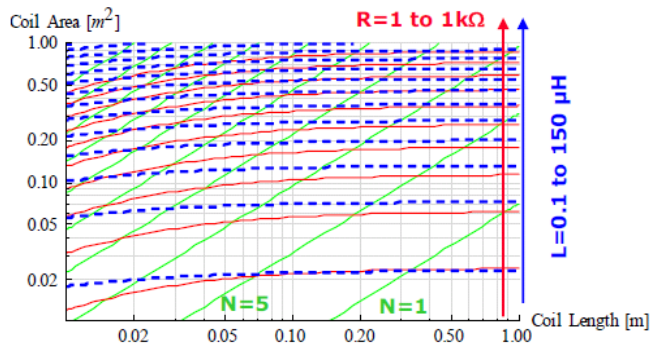

**Fig. 7:** Chart showing coil parameters $L_C$ (in blue), $R_C$ (red) and $N_T$ (green) as a function of coil length and total area, assuming $w_c$ = 10 mm and $\varnothing_w$ = 0.06 mm

## 4 Coil winding methods

Once the general parameters of a coil such as geometry, number of turns etc. have been chosen, then different manufacturing methods can be employed, each one with specific strengths and weaknesses. The main possibilities are treated in the following sections.

### 4.1 Single-strand coils

In the simplest case, a single strand of wire can be wound around the form in a helical pattern to make a coil. This basic technique is widely adopted in the larger context of discrete electronic inductors and is implemented in a number of commercial winding machines, all based on a simple combination of rotational plus cyclic translational motion of the winding form w.r.t. the wire bobbin. This method, clearly well-suited to the semi-automatic production of large numbers of identical parts, has two main disadvantages: 1) the maximum size allowed for the coil is limited by the size of the winding machine; and 2) the geometrical quality of the resulting coil can be quite bad, especially for high turn counts as shown in Fig. 8. In our own experience, coils up to 3000 turns fabricated with this method are used almost exclusively for applications such as quench localization in superconducting magnets, where the accuracy of the geometrical calibration factors (see Section 7.1) is of secondary importance. Wire in gauges from 32 to 200 micrometres is commonly used. Wire tension should be monitored accurately during winding, because the wire (especially the smaller one) breaks very easily.

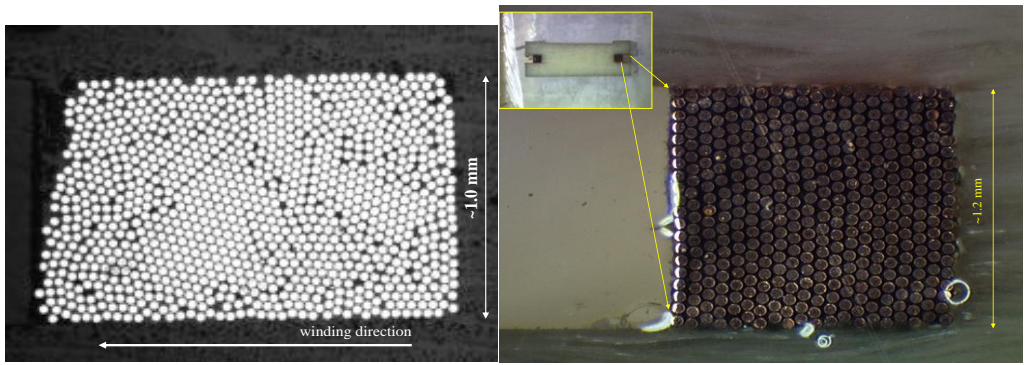

**Fig. 8:** Comparison between the cross-sections of a 900-turn single-filament, machine-wound (left) and a 400-turn, 20-filament manually wound coil (right)

Figure 5 shows a number of samples of single-strand coils taken from CERN stock. To enhance long-term stability after winding we brush on an outer layer of methyl acetate glue (Cementit™ Universal) diluted with acetone. In case of short circuits or other damage, it is possible to strip off the wire and recover the form by dissolving the glue with more acetone or by heating it up to around 100°C.

### 4.2 Litz-wire coils

The winding process can be greatly accelerated by using pre-assembled cables made with a plurality of wires. The traditional solution consists in the so-called 'Litz' wire (from *litzendraht*, 'woven wire'), a kind of cable including up to several thousand extremely fine insulated filaments (see Fig. 9). Litz wire has been used extensively since the 1940s in the radio frequency domain, where the subdivision of the conductor substantially reduces the AC losses due to skin effect. By using a wire with the appropriate number of conductors a whole coil can be made with just one winding turn, which is a big advantage in case of awkward geometries (e.g., very long coils, or thin grooves carved in a shaft). Unfortunately, this fabrication method carries some heavy penalties:

– there must be as many solderings as there are turns, and the wires to be connected must be painstakingly singled out one at a time;

- more importantly, each conductor follows a spiraling path along the coil and thus its transverse position undulates with a period equal to the twist pitch of the cable, and an amplitude up to its diameter. This implies that the local coil coefficients may vary in the range of several per cent, so that accurate coefficients can be obtained only by averaging over a given longitudinal tract.

At CERN, the most prominent use of Litz wire can be found in the coils for integral measurement of main Super Proton Synchrotron dipoles, which date back to the 1980s and have unique length requirements (more than 7 m).

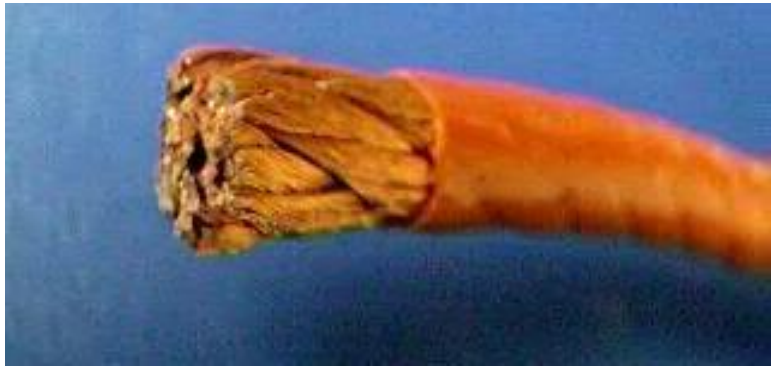

**Fig. 9:** An example of multi-filament Litz wire [from www.litz-wire.com].

### 4.3  Multi-strand coils

A better performing alternative to Litz wire is the multi-strand wire, a pre-formed ribbon which allows precise layering over multiple turns leading to very uniform geometries and highly accurate measurements. Examples of application can be found, e.g., at SLAC [16] and at CERN. The cable used at CERN is manufactured practically on-demand by MWS Wire Industries, Westlake Village, CA and consists of from 3 up to 20 polyurethane-insulated filaments with diameter from 170 μm down to 60 μm respectively. The filaments are bonded with polyvinyl butyral (PVB), a compound easily soluble in alcohol which facilitates soldering of the connections.

Winding a multi-strand coil, as shown in Fig. 10, is a fully manual procedure that requires skill and care. The winding form, which for practical reasons has in our case a maximum length of about two metres, pivots around a horizontal axle while the wire is pulled taut and guided to slide into the groove. As the wire rolls off the spool it dips in a bath of Araldite™ AY103 with HY991 hardener (a transparent, low-viscosity, general-purpose epoxy adhesive). After winding, the coil is inserted into a polymerization clamp to squeeze air bubbles out and to avoid the possible bulging of the wire at the corners (note that the wire tension tends to drop naturally along the short edge). A release agent (e.g., Teflon spray) should be added at this stage to prevent the coil from sticking to the clamp. The curing is carried out under IR lighting for a few hours at a temperature generally below 100°C, a precise thermal cycling being not strictly necessary. As a certain amount of springback upon removal of the clamp is to be expected, the geometry of the finished coil must be checked accurately.

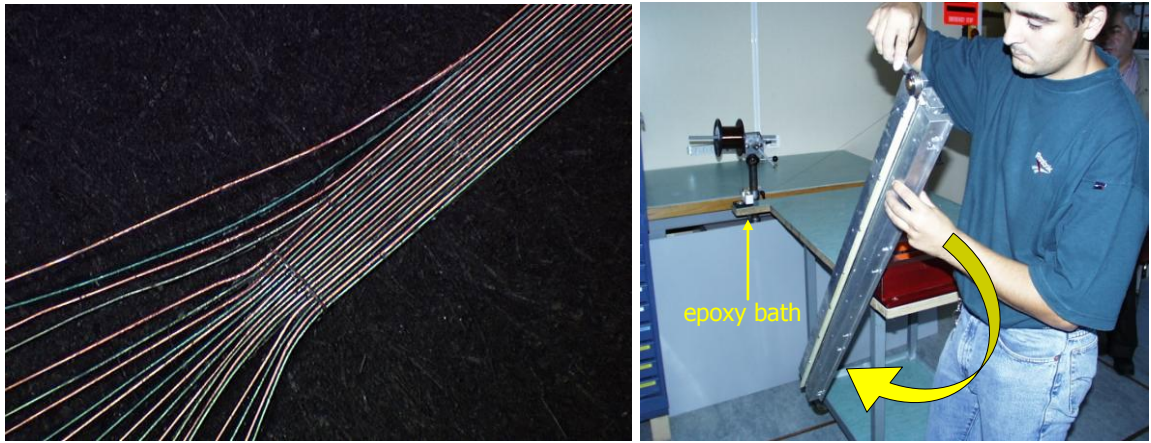

**Fig. 10:** An example of 20-filament flat wire (left) and a coil being manually wound (right)

The benefits deriving from this winding method are evident in Fig. 8, which shows the beautifully regular cross-section of a 20×20 turns coil developed to measure superconducting LHC magnets at room temperature (hence with currents of a few ampere and fields of a few millitesla). In our experience, such a high turn-count is unfortunately rather hard to obtain as a large proportion of the units (up to 2/3) is wasted, mostly due to the thin wire breaking during winding or the insulation tearing under the pressure of the clamps.

The electrical connections, very delicate due to the small size of the wires, must present the smallest possible area to the changing flux so as not to affect accuracy (this is important if the coil is designed to be calibrated and/or used when fully immersed in a magnetic field, see Section 7.2.1). First of all, the two ends of the ribbon have to be folded and overlapped as shown in Fig. 11. Then, all necessary solderings are made under a binocular microscope with the help of a tiny PCB connector, made with a complex procedure that involves an initial machining phase followed by aluminization and then deburring with a paste made by mixing micro glass balls, cementite and acetone. The connector should be glued or built into the coil to be easily removable seeing that, as one might expect, short circuits or broken wires are quite common. A thermo-retractable sleeve wrapped around the connector and the output cable is mandatory to relieve stresses and improve reliability. To conclude, it should be noted that this type of connector represents a constraint w.r.t. the minimum transversal size of a coil, and this may limit the applicability of this technology to very-small-aperture magnets.

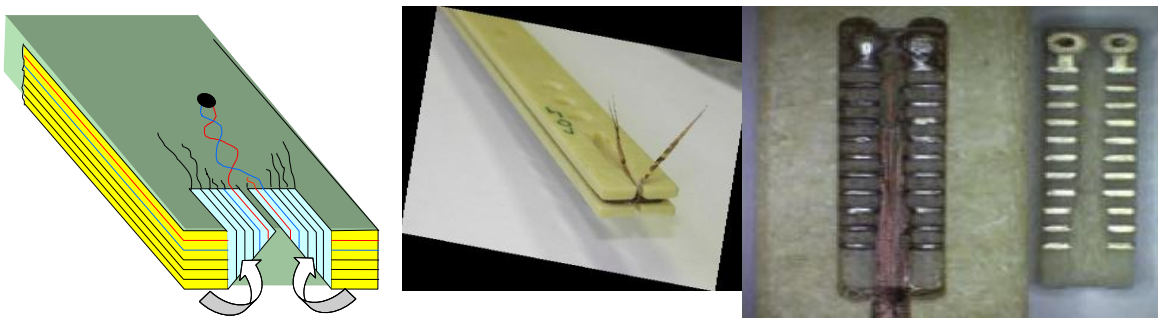

**Fig. 11:** Detail of the electrical connections of a multi-filament ribbon. The drawing on the left shows schematically the folding given to the ends of the wire (for clarity only one turn is represented). In reality, the folds are overlaid to minimize flux change pick-up (centre). The photo on the right shows the 10-mm PCB used to make the connections.

## 4.4 Printed Circuit Board coils

A rather different technological option for building coils consists in the possibility of laying spiral traces on a PCB, as shown in Fig. 12 [17]. A coil built with this technique is characterized by very precise (few micrometres) but relatively sparse traces, with a width of the order of 100 μm, thickness about 30 μm, and horizontal gap between traces also about 30 μm. The figure shows a board designed to be a part of a multi-layer assembly, as one can see from the many small metalized holes included to connect the 'turns' between layers (assemblies with up to 32 layers have been realized, although the cost tends to escalate rapidly).

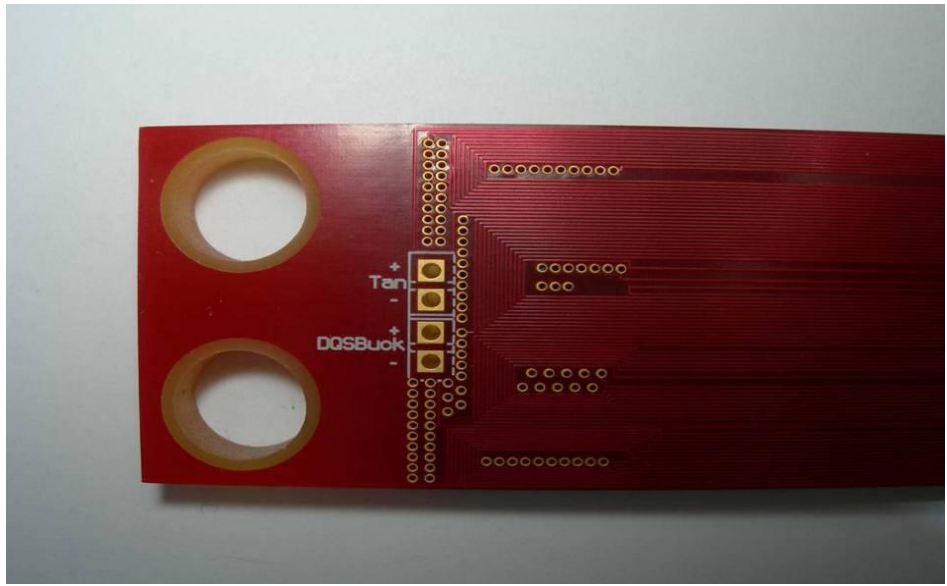

**Fig. 12:** Quadrupole-bucked, multi-layer PCB coil (courtesy of J. Di Marco, Fermilab).
The large holes on the left are used for alignment.

This kind of coil has some key advantages such as the possibility of industrial (read cheap and reproducible) manufacturing of large runs and the potential to scale down to coil sizes of a few millimetres, very difficult to make otherwise. Moreover, one may take advantage from the very precise positioning of the tracks to implement advanced multipole compensation schemes, e.g., dipole or quadrupole compensation with very large bucking factors (in the 1000–10 000 range) or more exotic higher-order schemes. On the other hand, there are a number of serious issues:

- owing to the low conductor density, the sensitivity is comparable with traditional designs only if many layers are stacked
- higher conductor density implies small cross-section and thus higher electrical resistance
- optimizing the trace layout is not a trivial task and requires close interaction with experts
- cost is high for short runs and multiple layers
- PCB lengths above ~1 m are not easily feasible
- the high aspect ratio of the coil must be taken into account in the calculation of coil sensitivity factors for accurate measurements [12]
- the end regions are comparatively large, so they too must be taken properly into account if the PCB is designed to be fully immersed in a magnetic field

The coil shown in Fig. 13 represents an interesting variation on this theme, where the rigid PCB is replaced by a flexible Kapton foil which is bent and glued directly onto the surface of a quartz rod [18]. While such a solution is undoubtedly elegant and precise, provided the foil is aligned correctly to the rod, the very low number of coil turns forces the instrument to turn rather fast.

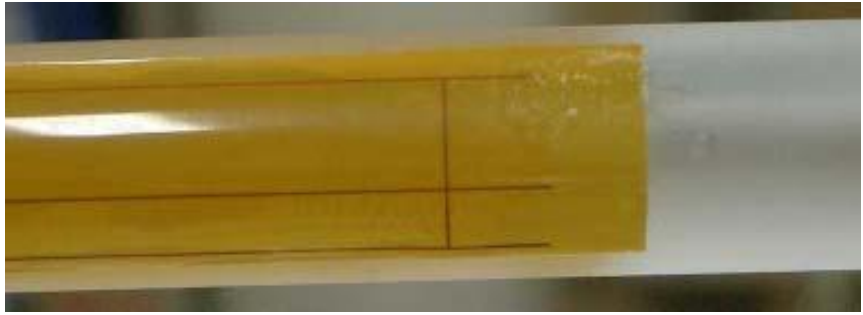

**Fig. 13:** A flexible coil printed on a Kapton foil and rolled around a ⌀20 mm rod (from Ref. [18])

### 4.5 Coils for special applications

Let us mention briefly a few special cases of application.

#### 4.5.1 Curved coils

For strongly curved magnets, accurate beam dynamics calculations may require the field integral to be measured along the actual path followed by the particles, which will be in general an arc of circle between two straight entry and exit segments. In such a case, the coil winding methods discussed above can be adapted at least in two ways:

– the coil is fabricated straight, then it is bent to the desired shape and fastened to a rigid support. The major drawback in this case is that bending stresses stretch the winding at the outboard and crush it at the inboard, causing the total surface to change and risking, at worst, detachment from the support at the inboard. The robustness and long-term stability of the coil are also impaired.
– The coil is wound directly on a curved form. While at the outboard the wire can be pulled taut as needed, at the inboard obviously this is not possible. The wire has to be glued to the form in a gradual and extremely careful way, possibly with the help of small clamps. Accurate geometry can be very difficult to achieve.

#### 4.5.2 Coils for cryogenic tests

Magnetic measurements and quench localization in superconducting magnets often require coils to operate at liquid helium temperature. Mainstream materials such as copper and glass-reinforced epoxy in general cope well with cryogenic conditions. Grade 3 wire should be specified to guard against thermal shocks, and differential thermal contractions (see Fig. 14) should be taken into account to estimate stresses in the windings:

$$\sigma_{Cu} = \frac{\int_{4.2}^{300}(\alpha_{Cu} - \alpha_{Support})dt}{\frac{\pi}{2}\frac{\varnothing_{Cu}^2}{A_{Support}}N_T\frac{E_{Cu}}{E_{Support}} - 1} E_{Cu} \qquad (7)$$

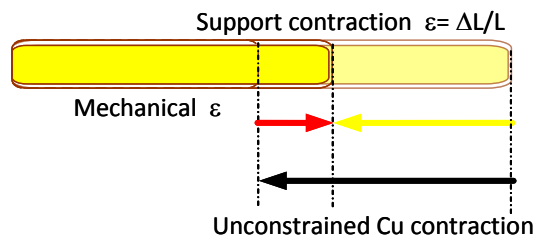

**Fig. 14:** Differential thermal contractions in a coil

Where necessary, specific adhesives such as Araldite™ GY285 (very low viscosity) + Jeffamine™ hardener should be used. This one, in particular, confers elasticity to the epoxy and the specific capacity to withstand differential expansion and thermal shocks without delamination.

*4.5.3   Coils for radiation environment*

In certain applications (e.g., 'B-train' field measurement system for real-time accelerator control) a coil has to be installed inside an operating magnet, with the consequent ionizing radiation risk due to synchrotron radiation and beam losses. Because of radiation, a search coil would suffer the same kind of effects which are well known for magnet coils, i.e.,

– copper can be activated if trace impurities are present (small diameter wire grades are normally 99.90% pure, with up to 0.04% oxygen to improve ductility);
– ceramic materials and glasses have very low activation;
– organic components like adhesives and cable insulation can be moderately activated. The major problem is related to certain polymers (e.g., polyethylene) which, in time, disintegrate completely.

In CERN's experience, over several decades no adverse effects have ever been observed in the few coils that have been exposed to radiation.

# 5   Coil head assemblies

Once the coil has been made, one has normally to fix it to an appropriate support to make a usable measurement head. While clearly there are as many possible ways to do this as there are different applications, we shall discuss here the rotating coil shafts made for LHC cryodipoles [19] as an example on which to base a few general considerations.

The coil shaft is shown in Fig. 15. The main points to be noted are the segmented structure, which allows the coil to follow the curved cold bore; the outer ceramic tubes, which guarantee the required torsional rigidity over the whole length of the shaft; and the vertical stack of three parallel, 1.1 m long rectangular coils within each segment, designed to provide dipole compensation with the third coil used for rotating mass symmetry and as a spare.

The major factor affecting the accuracy of multipole measurements is the alignment of the coil w.r.t. the rotational axis of the assembly, as even small errors in either radial position or angular orientation dramatically affect the harmonics of interest. In our case, the alignment is provided by a number of ceramic pins, while the fastening function is performed with nylon screws and secured by applying a line of glue along the edges of the top and bottom coils. The adhesive used is UV Loctite™ 322, which makes a flexible bond that can be scratched off if the segment has to be disassembled for repairs. On the downside, this glue has a very low viscosity and tends to penetrate between the coil and the support, so UV light must be applied immediately after deposition to accelerate the polymerization and prevent any possible swelling beneath the coil. Another caveat concerns poor long-term behaviour, especially when repeated thermal cycles or shocks are applied, as the glue tends to crumble and the residue can pollute the ball bearings.

An important component in this shaft is represented by the flexible bellows connecting adjacent segments, which allow quasi-homokinetic rotation in the bent configuration. The bellows are made of titanium, which has both the large elastic range and high electrical resistivity required by its function. It should be noted that the bellows are fastened to the ceramic shaft by means of Araldite. As the bond between the titanium and the ceramic was initially found not to be good, the surface was first roughened with acid (unsuccessfully) and then etched mechanically.

The last components to be highlighted are the ball bearings, which must be rigorously non-magnetic and non-conducting in order to work inside a 9 T field. The parts chosen are made of $Si_3N_4$ and are characterised by extremely low friction and smooth movement. They are mounted permanently at the extremity of each shaft segment, so as to provide a stable axis reference for calibration. While many suppliers of standard sizes exist, unfortunately our application required a custom design that could be obtained only from one manufacturer (Cerobear GMBH, Herzogenrath, DE) which, predictably, meant high costs.

A very interesting alternative is represented by the so-called Olive Hole Ring Jewel bearings, as used for instance at SLAC [11]. These parts are made of exceptionally hard artificial sapphire (monocristalline $Al_2O_3$) and offer a very small static friction coefficient of 0.15 associated to an extremely low price. Their major drawback lies in the fact that they cannot bear any axial load, so existing mechanical designs have to be modified accordingly.

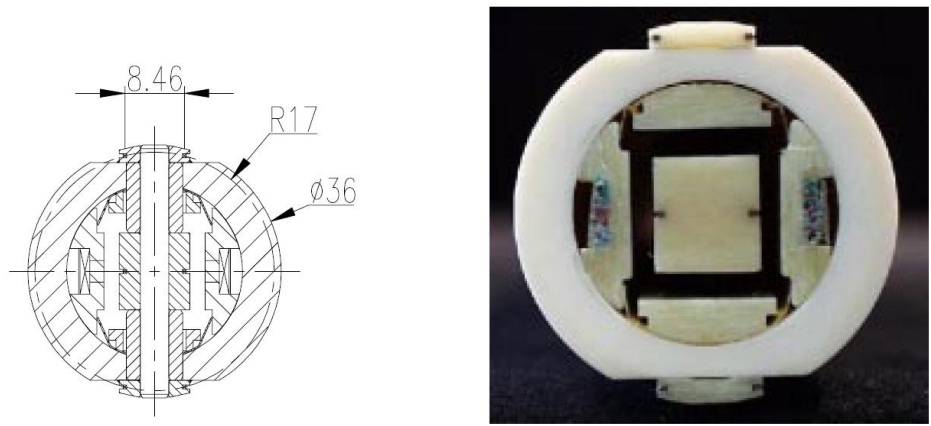

**Fig. 15:** Cross-section of a 15 m long, dipole compensated coil shaft for LHC main cryomagnets

# 6  Quality controls

At various stages during the fabrication of a search coil it is possible to carry out a number of quality-control checks meant to verify conformity with specifications. Simple go/no go tests and more precise measurements have to be done between manufacturing steps, prior to delivery, and also during operation (routine or on-call maintenance). In the following sections we shall discuss different kinds of checks as they are carried out at CERN.

## 6.1  Visual and geometrical inspection

To start with, careful visual inspection can easily reveal irregularities in the geometry as well as air bubbles, cracks or swellings in the epoxy (a binocular microscope can be of help). Next, accurate mechanical measurements of the length and width of both winding form and finished coil are mandatory to assess the quality of the winding and curing process, as well as to gather essential parameters for the calculation of calibration coefficients (see Section 7.1). While sub-micrometre 3D coordinate measuring machines are nowadays readily available and provided with knife-edge feelers that can penetrate easily into the winding groove, reasonably good results can be obtained inexpensively by means of simple rigs such as that shown in Fig. 16. In this case, a standard mechanical micrometer is made to slide along all four edges of the coil, which is then reversed to repeat the length and width measurements and allow cancellation of systematic offsets by averaging.

The data obtained from tests done on a batch of 15 1150 mm × 8.5 mm, 6 × 6 turn coils show that longitudinal variations of coil width amount to about 1% and are in good measure due to the irregularity of the winding support. Local variations of the winding width amount to about 10% and add considerably to the total spread, despite the comparatively high quality obtained with 6-strand flat cable. The main source of this variation seems to be the thickness of the polymerized epoxy layers, while wire diameter variations play a minor role. In addition, comparison between the mechanical coil area thus obtained and the magnetic surface (Section 8.1) for a more recent series of about 150 units shows that the spread of the correlation is about 1%, i.e., adequate to spot and reject major faults but not good enough to allow accurate prediction of magnetic surface from mechanical measurements.

For the sake of completeness it should be remembered that accurate geometrical measurements are in principle possible with microphotogrammetric techniques, both optical and radiographic (in this case, non-destructive).

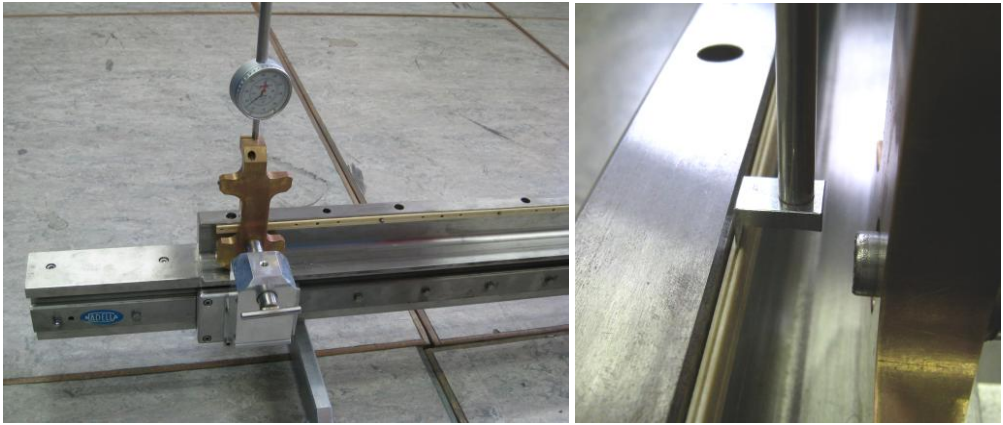

**Fig. 16:** A mechanical calibration rig for the geometry of rectangular coils. The micrometer on the left can be turned by 180° to cancel out offsets and by ±90° to measure the overall length. The detail on the right shows the knife-edge feeler used to get into the winding groove (typically 0.4 to 1.2 mm wide).

### 6.2 Magnetic measurement of coil width

As the width of a coil is of paramount importance for accurate measurements (see Section 7.2) we shall now describe a measurement method which is based on a strictly localized magnetic field and is therefore able to provide information on its magnetic equivalent directly. The principle is illustrated in Fig. 17, which shows the field profile we would ideally like to obtain along a given coil, corresponding to the total flux:

$$\Phi = N_T \int_0^L w(s)B(s)ds \approx N_T \int_{\bar{s}-\Delta/2}^{\bar{s}+\Delta/2} w(s)B(s)ds \approx N_T w(\bar{s}) \underbrace{\int_{\bar{s}-\Delta/2}^{\bar{s}+\Delta/2} B(s)ds}_{Bd\ell_{ref}}. \tag{8}$$

Once the flux is obtained with any one of the standard methods (i.e., by flipping or translating the coil, by pulsing or by AC modulating the field) the average equivalent magnetic length can be easily calculated from

$$w(\bar{s}) = \frac{\Phi}{N_T Bd\ell_{ref}}. \tag{9}$$

A possible way of obtaining the desired flux distribution based on permanent magnets and a suitably shaped iron yoke is also shown in Fig. 17. Alternative implementations based, for example, on AC coils may provide higher sensitivities, although permanent magnets ensure the stability required for accurate results. In all cases, the iron yoke has the important role of carrying the flux return away from the coil, lest the net flux measured drop dramatically (see the dotted curve in the figure).

It should be noted that, unless one knows with extreme precision the field map of the magnetic source and thus the absolute value of $Bd\ell_{ref}$ [see Eq. (16)], the width measurement obtained has mainly a relative character. The absolute value can be recovered by comparing the profile $w(s)$ to an integral measurement providing the missing average value (see Section 7.2).

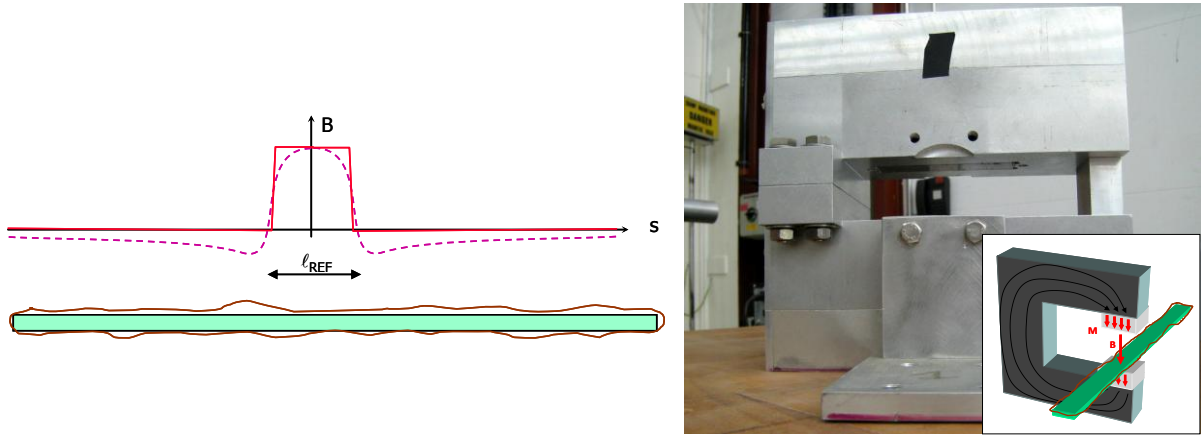

**Fig. 17:** Magnetic measurement of local coil width. The ideal and actual profile of the test field are shown on the left. The photo of an existing field source and its schematics (inset) are shown on the right.

## 6.3 Electrical quality controls

Basic electrical continuity tests should be repeated frequently throughout all phases of manufacturing, so as to detect as early as possible units with a broken wire. Once the coil is finished, its resistance should be measured as precisely as possible (i.e., with the four-wire method) in order to check for possible short circuits and to allow for later correction of the errors caused by finite input impedances (see Section 2.2). Referring back to Eq. (6-2), since the resistance is proportional to the number of turns we obtain

$$\frac{1}{R_c}\frac{\partial R_c}{\partial N_T} = \frac{1}{N_T} \quad . \tag{10}$$

One can see that in case of high turn count the sensitivity of the measurement drops dramatically: e.g., for a 400 turn coil the short of a single turn lowers the resistance by only 0.25%, which is quite difficult to measure correctly (consider that such a variation for copper is equivalent to a temperature uncertainty of just 0.5°C). The coil self-inductance given by Eq. (6-3), on the other hand, is purely a function of geometry and therefore is less prone to ambient perturbations. As the dependence upon the number of turns is squared, the sensitivity to shorts will be inherently higher[2]:

$$\frac{1}{L_c}\frac{\partial L_c}{\partial N_T} = \frac{2}{N_T} \quad . \tag{11}$$

---

[2] In fact, the shorted turn may add a mutual inductance that further subtracts from the initial self-inductance.

## 6.4 Coil polarity check

The polarity of a coil, i.e., the left- or right-handedness of the winding, is a seemingly simple issue that causes many headaches in practical situations, especially in complex acquisition systems where the signal path includes perhaps dozens of connections and therefore possibilities for inversions. The device shown in Fig. 18, based on a pair of AC-driven coils (of known polarity) represents an inexpensive way of resolving the ambiguity by using an oscilloscope to compare the phase of the source to that of the coil output. The optimal frequency range is typically around a few kHz, i.e., high enough to provide a strong output signal, yet lower than possible resonances.

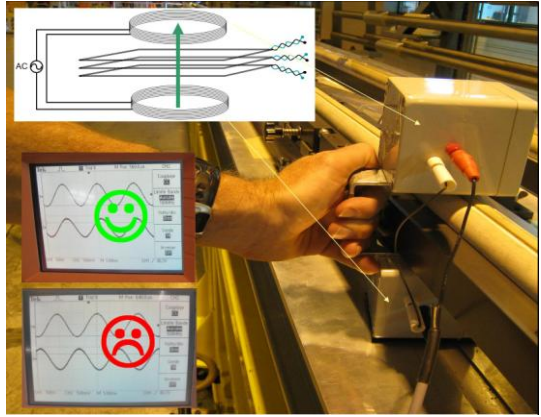

**Fig. 18:** AC coil polarity checker, testing one of the long LHC coil shafts

## 6.5 Magnetic quality controls

As one might expect, the surest way to prove that a coil is fit for magnetic measurements is indeed to go and actually make some, possibly in a well-known reference magnet which can be used to compare measured field and harmonics with expected values. Moreover, visualization and analysis of coil voltages, fluxes and other signals in the data acquisition and processing chain can often provide clues suggesting the source of observed malfunctions. In Fig. 19 we can see an example, taken from the test campaign of LHC dipoles, in which measurement results were intermittently grossly inaccurate while static resistance and polarity checks showed no apparent fault in the coils. In fact, the yellow absolute voltage trace shows clearly a disturbance occurring systematically at a certain angle during the forward rotation. This might be attributed to a failing contact, which was indeed identified and repaired.

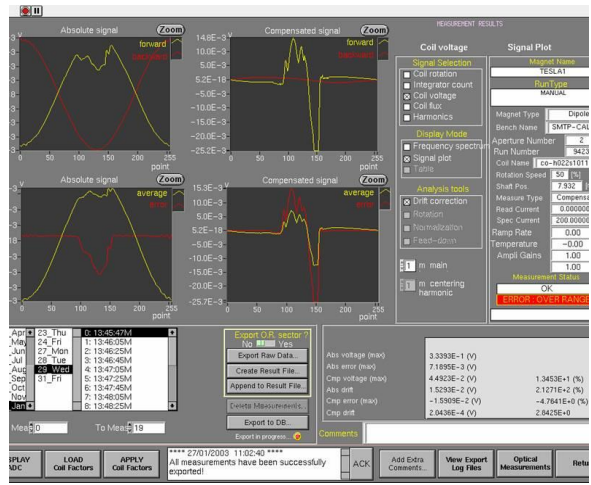

**Fig. 19:** Example of on-line detection of a failing electrical connection from the voltage signals of a rotating coil

## 7 Calibration of search coils

### 7.1 Basic theory of calibration

We shall henceforth deal with the principle and practice of coil calibration, referring essentially to the theory of harmonic coil measurements as developed in other lectures of this course [10], [20]. To recapitulate on the essentials: in the so-called thin-lens approximation, valid when the magnet length is much smaller than the wavelength of betatron oscillations, the field can be adequately described by a 2D power series expansion with complex coefficients $C_n = B_n + iA_n$, i.e., the harmonics. The flux $\Phi$ seen by a rotating search coil as a function of the azimuth $\vartheta$ is given by

$$\Phi(\vartheta) = \Re\left(\sum_{n=1}^{\infty} \frac{\kappa_n}{r_{ref}^{n-1}} \mathbf{C}_n e^{in\vartheta}\right) , \qquad (12)$$

where the $\kappa_n$ are complex coefficients known as coil sensitivity factors depending upon the geometry of the coil. In the case of an ideal rectangular shape[3] they can be defined as

$$\kappa_n = \frac{N_T L}{n}(z_2^n - z_1^n) = \frac{N_T L}{n}\left(R_2^n e^{in\frac{\alpha}{2}} - R_1^n e^{-in\frac{\alpha}{2}}\right)e^{in\varphi_0} , \qquad (13)$$

where $z_1 = R_1 e^{i(\varphi_0-\alpha/2)}$ and $z_2 = R_2 e^{i(\varphi_0+\alpha/2)}$ represent the position of the coil windings in the complex plane (x,y) at the start of the voltage integration, $\varphi_0$ is the initial phase, and $\alpha$ the coil aperture. By measuring flux increments over a discrete number $N$ of angular steps and computing their Fourier transform coefficients $\Psi_n$, the field harmonics can be reconstructed from

$$\mathbf{C}_n = \frac{2 r_{ref}^{n-1}}{N} \frac{\Psi_{n+1}}{\kappa_n}, \quad n = 1 .. \frac{N}{2} . \qquad (14)$$

On the basis of this relationship between coil and field coefficients, the process of coil calibration can be defined as **finding the set of $\kappa_n$ needed to infer field harmonics $C_n$ from the measurement of $\Phi(\vartheta)$.**

The expressions for the lowest-order $\kappa_n$ in three common cases are written out in full in Table 3. All coefficients can be calculated from the coil length $L$, width $w$, and average radius $R_0$; they all are proportional to total coil area $A_c = N_T L w$; they all increase like $R_0^{n-1}$. It is worth while pointing out that the $\kappa_n$ are independent from the field being measured, i.e., search coils are inherently linear sensors. Indeed, contrary to a bafflingly popular misconception, the measurement gets easier as the field gets stronger as the S/N ratio of the output voltage improves.

---

[3] The correction to be applied to take into account the finite size of the winding, as described in [12], is completely negligible for dipole, $<10^{-4}$ for quadrupole, and $<10^{-3}$ up to dodecapole *if* the cross-section is square and smaller than 1 mm$^2$, and the rotation radius is larger than 10 mm.

**Table 3:** Lowest-order coil sensitivity factors in three common cases

| n | Radial coil $\varphi_0 = 0$ | Tangential coil $\varphi_0 = \pi/2,\ w = 2R_0\sin\alpha/2$ | Tangential coil $\varphi_0 = \pi/2,\ \alpha \approx 0$ |
|---|---|---|---|
| $\kappa_1$ | $N_T L w$ | $N_T L w$ | $N_T L w$ |
| $\kappa_2$ | $\frac{1}{2} N_T L w R_0$ | $i \cos\frac{\alpha}{2} N_T L w R_0$ | $i N_T L w R_0$ |
| $\kappa_3$ | $N_T L w \left(\frac{w^2}{12} + R_0^2\right)$ | $-\frac{1}{3}(1 + 2\cos\alpha) N_T L w R_0^2$ | $-N_T L w R_0^2$ |
| $\kappa_4$ | $N_T L w R_0 \left(\frac{w^2}{4} + R_0^2\right)$ | $-i \cos\alpha \cos\frac{\alpha}{2} N_T L w R_0^3$ | $-i N_T L w R_0^3$ |
| $\kappa_5$ | $N_T L w \left(\frac{w^4}{80} + \frac{w^2 R_0^2}{2} + R_0^4\right)$ | $\frac{1}{5}(1 + 2\cos\alpha + 2\cos 2\alpha) N_T L w R_0^4$ | $N_T L w R_0^4$ |
| $\kappa_6$ | $N_T L w R_0 \left(\frac{w^4}{16} + \frac{5}{6} w^2 R_0^2 + R_0^4\right)$ | $\frac{i}{3} N_T \cos\frac{\alpha}{2}(4\cos^2\alpha - 1) L w R_0^5$ | $i N_T L w R_0^5$ |

When we are faced with the practical problem of finding the $\kappa_n$ of a coil, different alternative routes can be followed. The dataflow diagram in Fig. 20 represents the relationships between the different variables and procedures entering into play. We see that

- The geometrical parameters needed to compute the $\kappa_n$ derive from a combination of nominal values, geometrical measurements and magnetic measurements.
- Purely geometrical measurements, such as those described in Section 6.1, can in principle provide all basic length and width parameters. In practical situations, however, only the length can be measured to the level of accuracy required for high-precision end results (typically $10^{-3} \sim 10^{-4}$). (NB: if the coil is meant to measure shorter magnets, including fringe fields, then total length and surface are irrelevant).
- Measurements in a suitable reference magnet, i.e., one whose coefficients $C_n$ are known, can be used to derive some of the $\kappa_n$ and then work backwards to infer surface, width, and radius, which are then used to compute the missing coil coefficients.

The magnetic calibration option is the one that gives best results, since all sources of error are included in the same way they will be in real use. Failing that, due for example to the lack of a suitable reference magnet, geometrical measurements remain the second-best alternative. Calibration measurements in a dipole (to obtain coil surface) and in a quadrupole (to obtain rotation radius) as described in Section 8 are the norm; using higher-order multipole magnets is far less common.

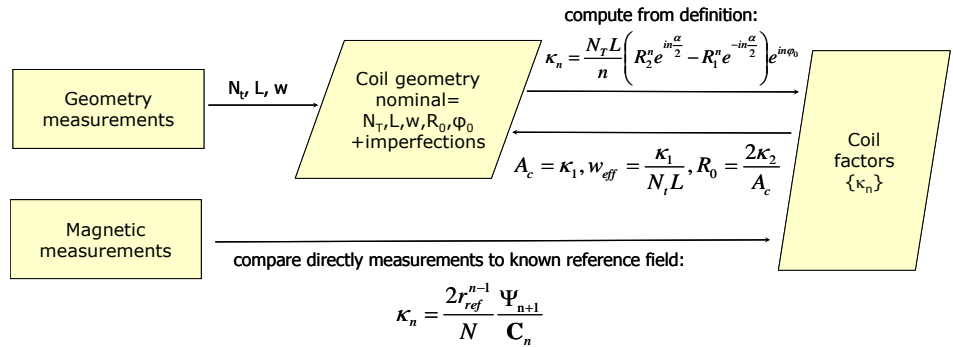

**Fig. 20:** Schematic representation of the dataflow for search coil calibration

## 7.2 Calibration of finite-length coils

In practical use, the theoretical considerations discussed above in the ideal 2D case must be extended to the third dimension, taking into consideration end effects linked to both the coil and the magnet. Let us take as an example the configuration shown schematically in Fig. 21, where we shall assume that the coil length $L$ is known accurately from geometrical measurements. We shall focus our attention on the case of a dipole magnet, the extension to other cases being trivial. The magnetic flux through the coil, which can be obtained, for example, by flipping the coil, fully rotating it, or keeping it fixed while the field is ramped up, is given by the following expression:

$$\Phi = N_T \int_0^L w(s)B(s)ds = \overline{B}\, \underbrace{\overline{w}_{eff} L}_{A_{eff}} = \overbrace{\overline{B} L}^{Bd\ell}\, \overline{w}_{eff}\ . \tag{15}$$

The main point here is that the quantity which is really of interest vis-à-vis accelerator physics is emphatically not this flux, but rather the integrated field (or $Bd\ell$, as it is commonly called), defined by (assuming the coil longer than the magnet):

$$\mathrm{B}d\ell = \int_0^L B(s)ds\ . \tag{16}$$

The following averaged quantities can also be defined:

$$\begin{cases} \overline{B} = \dfrac{1}{L}\int_0^L B(s)ds & \text{Average field} \\[2mm] \overline{w}_{eff} = \dfrac{\int_0^L N_T w(s)B(s)ds}{\int_0^L B(s)ds} = \dfrac{\Phi}{\mathrm{B}d\ell} & \text{Average effective width} \\[2mm] A_{eff} = L\overline{w}_{eff} & \text{Effective coil surface} \end{cases} \tag{17}$$

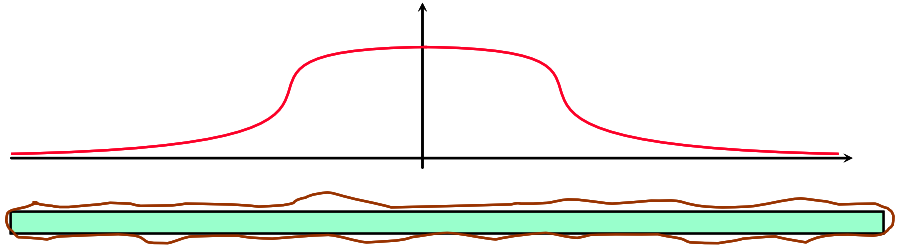

**Fig. 21:** Schematic representation of longitudinal field profile along a coil

In the general case, it is important to note that unless the width is strictly constant along the coil, the magnetic flux cannot be obtained from the mere knowledge of average field and average width alone, since

$$\tfrac{1}{L}\int_0^L w(s)B(s)ds \neq \tfrac{1}{L}\int_0^L B(s)ds\, \tfrac{1}{L}\int_0^L w(s)ds\ . \tag{18}$$

The effective width over the profile of interest [Eq. (17-1)], in fact an average width weighted with the field and lumped together with the number of turns for convenience, turns out to be the key parameter. Knowing that we can easily derive the wanted field properties from the flux measurement

$$\begin{cases} \overline{B} = \dfrac{\Phi}{A_{eff}} = \dfrac{\Phi}{L\overline{w}_{eff}} = \dfrac{Bd\ell}{L} \\ Bd\ell = \dfrac{\Phi}{\overline{w}_{eff}} \end{cases} \quad . \tag{19}$$

The central problem of calibration lies therefore in the correct evaluation of the effective width, which requires the same $B(s)$ profile during calibration and normal measurements. Failing that, even in the most favourable case, i.e., $B$ = const., the error committed on the field integral will be of order $\delta w/w$, easily in the per cent range for narrow coils. Two extreme scenarios can be discussed:

**a) local measurement: $L_c \ll L_M$**

assuming that longitudinal and transversal variations across the coil of $B$ are negligible, the simple arithmetic average of the width is sufficient. The calibrated value $A_{eff}$ can be used to obtain the average field from Eq. (19).

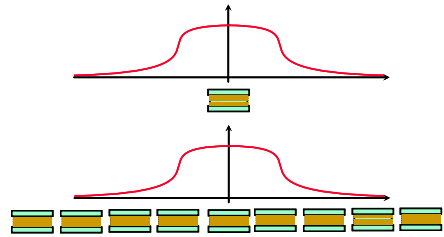

Utilization:
– local field quality (e.g. diagnostic tool for long superconducting magnets),
– integral measurement by scanning (errors tend to vanish as the coil gets shorter).

**b) integral measurement: $L_c > L_M$**

even perfect knowledge of $w(s)$ does not
permit accurate absolute measurement of $Bd\ell$

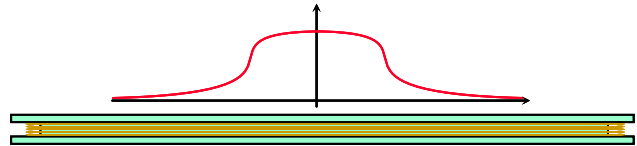

Utilisation:
– relative measurements: e.g., tracking a reference magnet, or measurement of normalized harmonic errors.

Concerning the role of absolute vs. relative measurements, it should be emphasized that the characterization of main synchrotron magnets requires simply that all magnets be equal to each other within a given tolerance, seeing that magnets are usually powered in series and no individual corrections are possible. Absolute field maps are mandatory, on the other hand, in the case of spectrometers, or for instance if precise beam energy matching of cascading accelerators is required. Where absolute calibration is really necessary, possible alternatives include:

- Establish the shape of the $B(s)$ profile with independent methods (e.g., low absolute accuracy Hall probe measurements, or scaling from known similar magnets) and work out the effective width from Eq. (17-2)
- Use an independent measurement system, e.g., a combination of NMR/Hall probes or, more effectively, a stretched wire system.

### *7.2.1    End effects*

Another factor to consider in order to obtain high accuracy measurements is the shape of the winding at the ends of the coil, which as we have seen in Section 4.3 must be somewhat rounded to ensure good bonding to the support. In case the coil ends stay outside of the magnetic field, including during calibration, the issue becomes obviously irrelevant. For simplicity, we shall consider advantages and drawbacks of two extreme alternatives:

1) Straight edges (rectangular coil)
   In this case, the coil length can be defined and measured precisely. This shape allows most easily the juxtaposition of several measurements to reconstruct the integral in long magnets. On the downside, very sharp corners are easily going to cut the wire during winding.

2) Semi-circular edges (racetrack coil)
   An accurate measurement of the average field becomes possible, in this case, only if the field distribution across the whole coil is very nearly uniform. Since the total flux can be expressed as follows (for a coil with total length $L+w$ and end radius $w/2$):

$$\Phi = N_T \int_{-w/2}^{0} 2\sqrt{\frac{w^2}{4} - s^2}\, B(s)ds + N_T w \int_{0}^{L} B(s)ds + N_T \int_{L}^{w/2} 2\sqrt{\frac{w^2}{4} - (s-L)^2}\, B(s)ds \ , \qquad (20)$$

easy calculation of $Bd\ell$ is possible only if the profile of $B(s)$ over the rounded regions is constant or at most linear.

## 7.3 Metrological considerations

We have seen that the calibration of a search coil from first principles (i.e., from the geometry alone) is not practical if high accuracy is sought, and the use of adequate reference magnets is therefore a necessity. This raises, inevitably, the question of how can we know the reference field in the first place. To frame the problem in the correct metrological context we have to consider first the relationships between the magnetic units for field and flux (T and Wb, or $Tm^2$ respectively) and the four base units for electromechanical problems (kg, m, s and A) in the International System (SI), as expressed by the diagram in Fig. 22 [21]. We find that:

1) **Magnetic flux [$\Phi$] = Wb**
   The weber is defined in terms of all four base SI units by means of a rather long chain of intermediate units:

$$Wb = V \times s = \frac{W}{A} s = \frac{J}{A} = \frac{N \times m}{A} = \frac{kg \times m^2}{A \times s^2} \ .$$

On account of this complexity, no primary standard exists, although the possibility of adopting as such the so-called flux quantum[4] $\Phi_0 = \frac{h}{2e}$ is being discussed and appears very likely in the long run [22]. While historically a number of secondary flux standards have been proposed and used (see for instance Hibbert's device [23]), nowadays the availability of ppm-level voltage and time references makes it far easier to calibrate the whole integration chain in terms of purely electrical quantities.

---

[4] The flux quantum is a universal constant, appealing because independent from any material properties, which covers an important role in the study of flux dynamics in superconductors. Since the flux quantum appears in the expressions for both the Josephson effect ($V = nf\Phi_0$) and the quantum Hall effect ($R = n\Phi_0/e$), it is being proposed as the basis for a future quantum redefinition of both the volt and the ohm, hence of all other SI units.

2) **Magnetic field [B] = T**

The field *B* (more accurately known as either magnetic induction or flux density, in order to distinguish it from the magnetic field *H*) can also be defined in terms of base SI units:

$$T = \frac{Wb}{m^2} = \frac{kg}{A \times s^2}.$$

Primary standards do not exist for the tesla, either. The accepted secondary standard is the NMR teslameter, an instrument based on the RF detection of nuclear resonance in a suitable sample that can routinely achieve absolute accuracies of a few ppm and resolutions of a fraction of a ppm over a wide range of measurement conditions [24]. This excellent performance is achieved by the transduction between field and frequency, which nowadays can be measured with extreme precision thanks to widely available high-precision digital oscillators. The main limitations of commercial units lie in the requirement of high field homogeneities (~10 ppm/mm, unless the gradient is somehow compensated) and low field rates of change ($10^{-2}$/s).

In everyday use, the availability of secondary standards of magnetic field and flux allows the practitioner to calibrate different types of field sources and instruments, which should be whenever possible cross-checked in redundant configurations to identify systematic errors and to increase the statistical confidence in the results.

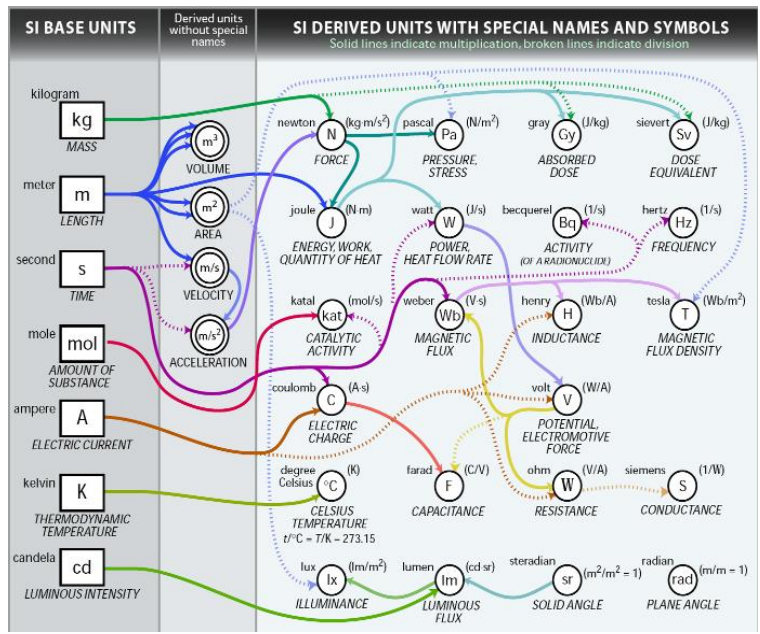

**Fig. 22:** Relationship diagram between base and derived SI units [from www.nist.gov]

### 7.3.1   *Calculable field sources*

While standard iron-dominated magnets are the most common instance of calibrated field sources, many issues linked to non-linear effects, ageing of the materials etc. add complications to their practical use. The so-called 'calculable' sources, on the other hand, attempt to realize secondary standards based as directly as possible on first principles, e.g., on well-known geometry, current and stable mechanics. Among the better-known sources of this kind we find the classical Helmholtz coils, shown in Fig. 23 [25].

In the simplest configuration, this source is made by a pair of identical and parallel circular coils separated by a distance equal to their radius *R*. As it can easily be proven, if both coils are fed the same current the generated field is very uniform since the lowest-order harmonic permitted by the symmetry (the sextupole) vanishes. Note that all even harmonic are also null, barring construction errors. The central field can be expressed as:

$$B_0 = \sqrt[3]{\frac{4}{5}} \frac{\mu_0 N_T I}{R} = 0.715 \frac{\mu_0 N_T I}{R} \tag{21}$$

and the relative field change $\Delta B/B_0$ is of the order of 6% at the centre of the two coils.

The main advantage of the Helmholtz arrangement is that the uniform field region is easily accessible. The concept can moreover be generalized to different configurations: for example, reversing the current in one if the coils create a good quality quadrupole field, while a 2D version of the device, made with rectangular coils, can be built by letting the coil width be √3 times the coil spacing. More complex arrangements with three or more coils are also possible to obtain better uniformity of the desired field, or possibly arbitrary orientations of the resulting field vector.

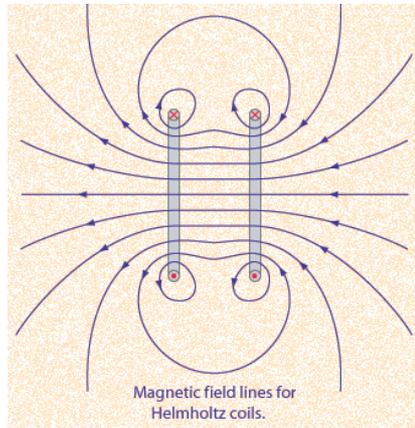

**Fig. 23:** Magnetic field produced by a standard Helmoltz coil configuration

## 8   Calibration of harmonic coils

In this section the specific techniques used to calibrate harmonic coils will be described in some detail. Referring back to Section 7.2 we shall deal, for the sake of simplicity, with the case of a short coil fully immersed in a magnetic field, i.e. we will consider all parameters averaged along the coil and ignore local fluctuations. All the methods illustrated here can of course be applied in the opposite limit, i.e., when the coil is shorter than the magnet; however, in this case the results must be understood to apply only to magnets having closely matching magnetic length and field profile w.r.t. the one used for calibration.

### 8.1   Surface

The total surface of a coil can be readily obtained by flipping it upside down in a uniform field of known average *B* over the coil area. Integration of the coil voltage gives

$$-\int_0^t V_c dt = \Phi - (-\Phi) = 2A_C \bar{B} . \tag{22}$$

This technique is fundamentally equivalent to a standard harmonic (rotating) coil measurement where only two integration points, 180° apart, are taken. The main advantage of the method lies in the fact that an integral w.r.t. time is transformed into an integral w.r.t. angular position, thus making the end result insensitive to speed and trajectory fluctuations. A second flip back to the original position is always advisable in order to estimate and correct the error due to the integrator drift, which will respectively add and subtract to the two results. Errors due to imperfect 180° rotation grow as (1-cos(angular error)) and are therefore often negligible, typically around $10^{-4}$ per 1°.

Figure 24 shows the reference dipole magnet installed in the CERN coil calibration laboratory, including the apparatus for precisely controlled coil flipping. The central part of the dipole is mapped to a high precision with an NMR probe, so that the average field across coils of different length can be easily calculated. Coils thus calibrated can be used to measure the integral of long magnets by means of longitudinal scanning. In such a case, there will be an inevitable error due to the fringe field having a gradient, hence the approximation $w_{eff} = w_{avg}$ does not hold anymore. This error can be estimated (and, being systematic, eliminated) by cross-checking the integral field with an independent method, for example a stretched wire. Typical accuracies of the order of a few $10^{-4}$ can be routinely achieved in this case.

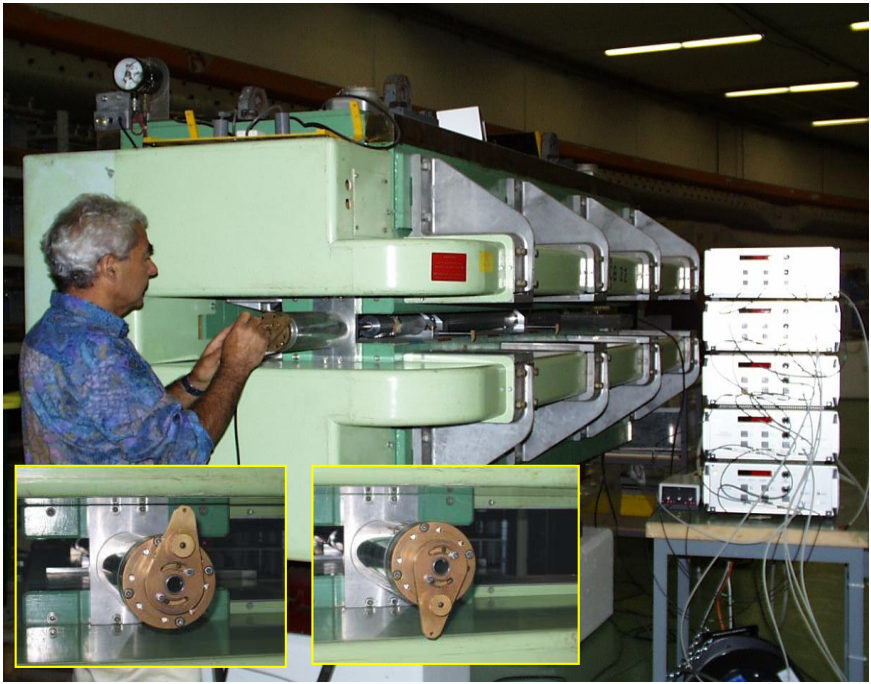

**Fig. 24:** CERN dipole apparatus for the calibration of coil surface and parallelism

## 8.2 Array parallelism

In many practical cases the measurement head includes an array of coils which are meant to have identical surface and orientation (see, for example, the compensated configurations described in Section 2.3, or the curved fluxmeter in Section 9). Any parallelism error between these coils is detrimental to accuracy as it may introduce spurious quadrature harmonics (e.g., a fictitious skew harmonic where the physical one is normal, and vice versa). The calibration of this error can also be carried out in a known dipole field, however, this time the coil must be flipped while *parallel* to the field.

With reference to Fig. 25, let us consider first the pair of coils 1 and 2, each having a tilt angle $\varepsilon$ w.r.t. the support and a surface $A$. The respective dipole sensitivity coefficients are

$$\kappa_1^1 = A_1(\cos\varepsilon_1 + i\sin\varepsilon_2) \approx A_1(1+i\varepsilon_2), \qquad \kappa_1^2 \approx A_2(1+i\varepsilon_1) \ . \tag{23}$$

And the dipole coefficient of the difference signal is

$$\kappa_1^{diff} = \kappa_1^2 - \kappa_1^1 \approx \Delta A + iA\Delta\varepsilon \; , \tag{24}$$

so the flux difference measured by flipping $\Delta\Phi$ can be used to derive the parallelism error $\Delta\varepsilon$:

$$\Delta\Phi = 2\Re\left(\kappa_1^{diff} B_1 e^{in\varphi_0}\right) = 2B_1\left(\Delta A\cos\varphi_0 - A\Delta\varepsilon\sin\varphi_0\right) \Rightarrow \Delta\varepsilon = \frac{\Delta\Phi}{2AB_1} \; . \tag{25}$$

If this angular difference exceeds the prescribed tolerance (usually a few milliradian), one should shim or file the support to compensate, and then iterate. By repeating the procedure it is clear that all coils can be made parallel to one selected as the master; however, the tilt of the master itself remains unknown. The additional procedure for the calibration of the absolute field direction is described in the next section.

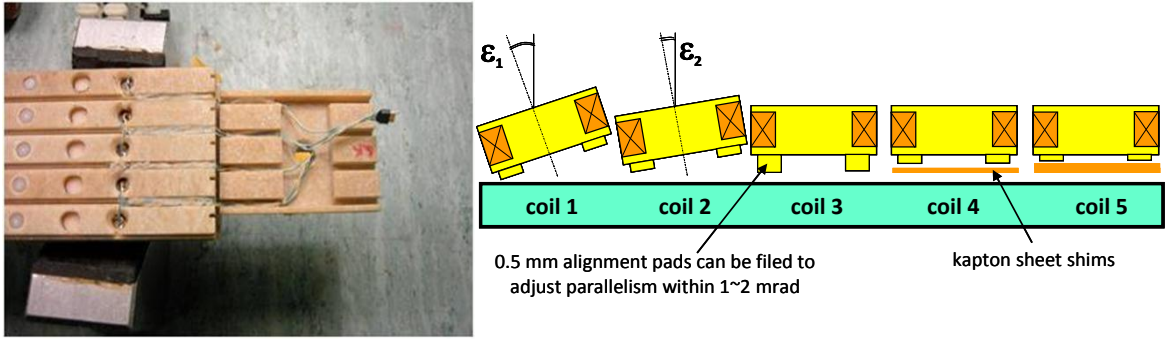

**Fig. 25:** A 5-coil array fluxmeter for quadrupole-compensated harmonic measurements, including a spare coil (left). Schematic representation of parallelism calibration procedure (right).

## 8.3 Field direction

The calibration of field direction measurement for a harmonic coil has the purpose of measuring the total angular offset $\Delta\alpha$ between the measuring coil and the angle corresponding to the start of coil voltage integration, usually marked by an index pulse from an angular encoder. The total offset shall include all contributions from different error sources, e.g., the angle between a coil and its support, the angle between support and encoder etc.

The simplest and most accurate method involves repeating a full harmonic measurement in a dipole magnet while reversing the relative orientation of coil and field, i.e., flipping either one by 180° around an axis normal to the coil rotation axis (see Fig. 26). Since the unknown offset respectively adds and subtracts to the measurement in the two different configurations, both the offset and the actual field direction $\alpha$ can be computed:

$$\begin{cases} \alpha_{meas}^1 = \alpha + \Delta\alpha \\ \alpha_{meas}^2 = \pi - \alpha + \Delta\alpha \end{cases} \Rightarrow \begin{cases} \alpha = \dfrac{\pi}{2} + \dfrac{\alpha_{meas}^1 - \alpha_{meas}^2}{2} \\ \Delta\alpha = -\dfrac{\pi}{2} + \dfrac{\alpha_{meas}^1 + \alpha_{meas}^2}{2} \end{cases} \tag{26}$$

When the flipping is not possible *in situ*, $\Delta\alpha$ may of course be measured in a convenient reference magnet and stored for later use. While the case of a dipole is somewhat more intuitive, the method can be extended to multipoles of every order. However, as the correct handling of the related inverse trigonometry in the analysis software is potentially messy, it is strongly suggested to double-check procedures and algorithms by applying known rotations to the magnet (if possible over 360°) in order to verify absolute values and signs of the measured rotations. Accuracies in the range of a fraction of a milliradian are routinely achieved with this method.

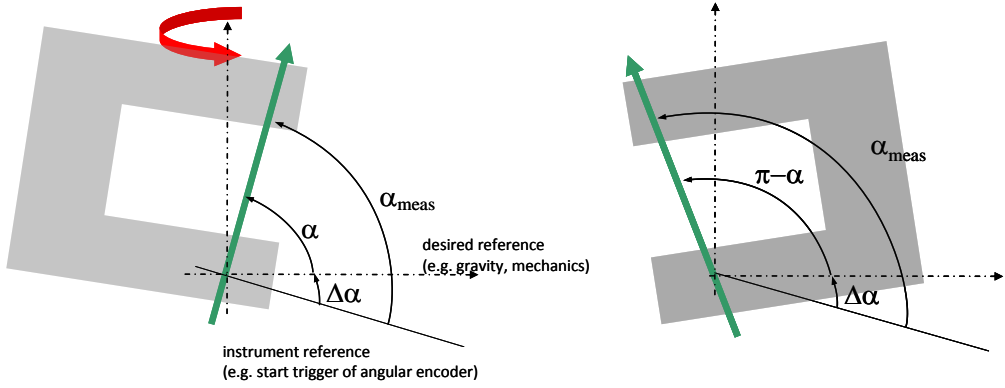

**Fig. 26:** Calibration of field direction in a reference dipole. The second configuration (right) is obtained by flipping either the instrument or the magnet by 180° around a vertical axis.

### 8.4 Rotation radius

As seen in Section 7.1, the measurement of all harmonics above dipole relies on accurate knowledge of the average coil rotation radius $R_0$. The calibration of this quantity is particularly difficult, not least because the mechanical position of the rotation axis must not change between calibration and regular use, as might well be the case if, for instance, different ball bearings are used.

We shall outline briefly a calibration procedure based on precisely controlled coil displacements inside a quadrupole magnet, as shown in Fig. 27. We sill assume the total coil surface $A_c$ to be known and, without loss of generality, an horizontal initial offset between rotation axis and magnetic centre. First, the quadrupole strength can be inferred from the flux change associated to a translation $\Delta x$:

$$B_2 = \frac{r_{ref}}{\Delta x} \frac{\Phi(0, x_0 + \Delta x) - \Phi(0, x_0)}{A_c} \ . \tag{27}$$

Note that, even in the case of a dedicated reference magnet, it is advisable to re-measure the strength upon every calibration campaign in order to avoid errors due to drifts (magnet and electronics), temperature, power supply instability, etc.

Assuming the quadrupole to be pure, the flux seen by the coil as a function of azimuth $\vartheta$ can be expressed as

$$\Phi(\vartheta) = \Re\left(\kappa_1 B_1 e^{in\vartheta} + \kappa_2 B_2 e^{2in\vartheta}\right) = \frac{A_c B_2}{r_{ref}}\left(x_0 \cos\vartheta + \frac{1}{2} R_0 \cos 2\vartheta\right), \tag{28}$$

where the $B_1$ term is generated by feed-down of the quadrupole. Among the many possible and essentially equivalent choices, we shall consider two coil rotations between $\vartheta = 0$, $\pi/2$ and $\pi$ to obtain

$$\begin{cases} \Phi(0) = \frac{A_c B_2}{r_{ref}}\left(x_0 + \frac{1}{2} R_0\right) \\ \Phi\left(\frac{\pi}{2}\right) = \frac{A_c B_2}{r_{ref}}\left(\phantom{-x_0} -\frac{1}{2} R_0\right) \\ \Phi(\pi) = \frac{A_c B_2}{r_{ref}}\left(-x_0 + \frac{1}{2} R_0\right) \end{cases} \Rightarrow \quad x_0 = r_{ref} \frac{\Phi(0) - \Phi(\pi)}{A_c B_2}, \ R_0 = r_{ref} \frac{\Phi(0) - \Phi\left(\frac{\pi}{2}\right)}{A_c B_2} - x_0 \ . \tag{29}$$

Analogous expressions can easily be derived for the *y* coordinate in the general case. For better results, it is suggested to centre the coil in an iterative fashion making use of the measured value of the offset. Once centred, the coil can be made to rotate over a full turn and four 90°-spaced flux increments can then be acquired. Since these should all be equal in absolute value, the average can be taken as the wanted result and the differences will give an estimate of systematic and random errors (as always, redundant measurements are the key to improved accuracy).

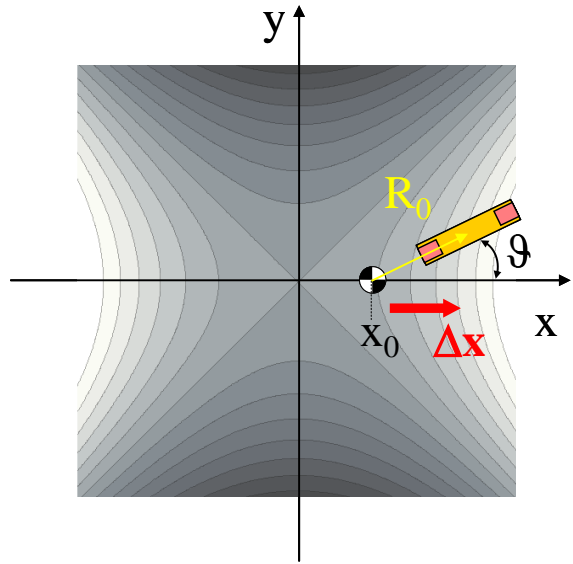

**Fig. 27:** Schematic representation of the calibration procedure for the average rotation radius of a harmonic coil ($R_0$) in a quadrupole magnet

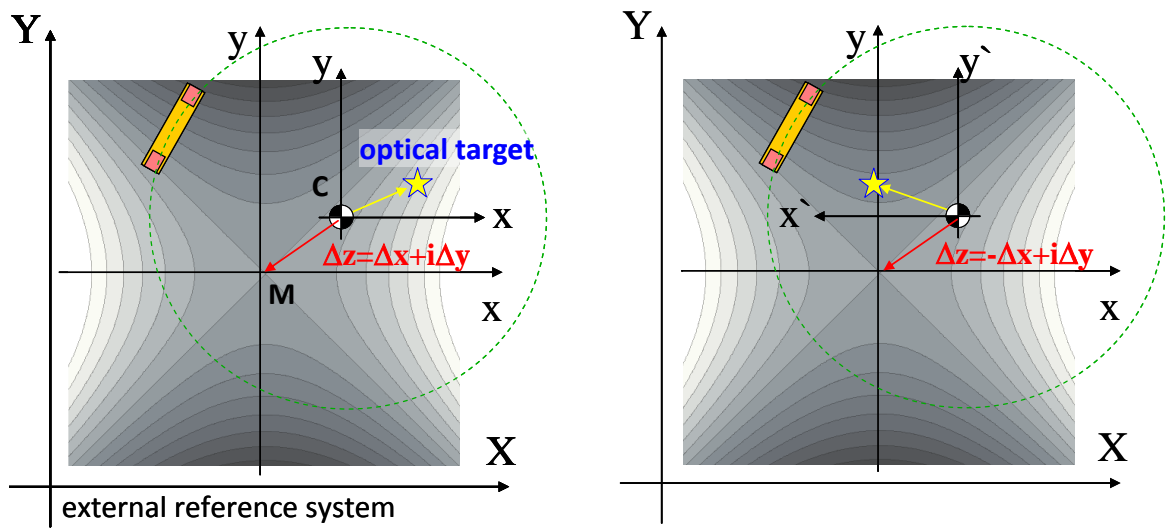

**Fig. 28:** Schematic representation of the calibration procedure for the magnetic centre using a harmonic coil in a quadrupole magnet. The second configuration (right) is obtained by flipping the coil by 180° around the vertical axis *y*. The optical target marked by the star is rigidly connected to the rotating coil.

## 8.5 Magnetic centre

Among the various existing methods for measuring the magnetic centre (or axis) of a multipole, the harmonic coil can provide excellent results as long as the position of the rotation axis is measured w.r.t. an external coordinate system, from which the centre can be then transferred to the magnet fiducial reference system. One of the possibilities consists in fixing an optical target on the rotation axis itself, so that it can be imaged by an external instrument such as a laser tracker or a theodolite.

The calibration of such a target can easily be accomplished in a quadrupole magnet, as shown in Fig. 28. We shall consider two configurations (1,2) in which the rotating coil probe is flipped by 180° around the vertical (*y*) axis. Let *(x,y)* be the reference axis of the coil, centred on its rotation axis C; *(X,Y)* the external coordinate system in which the optical target T is measured; $\Delta z = \Delta x + i\Delta y$ the complex offset of the magnetic centre M w.r.t. C, as obtained by feed-down from each measurement. The horizontal components of the centre position ($X_M$) and optical target offset ($x_T$) can be obtained as follows:

$$\begin{cases} X_{T1} = X_M - \Delta x_1 + x_T \\ X_{T2} = X_M + \Delta x_2 - x_T \end{cases} \Rightarrow \begin{cases} X_M = \dfrac{X_{T1} + X_{T2}}{2} + \dfrac{\Delta x_1 - \Delta x_2}{2} \\ x_T = \dfrac{X_{T1} - X_{T2}}{2} + \dfrac{\Delta x_1 + \Delta x_2}{2} \end{cases}. \tag{30}$$

The vertical component can be worked out in a similar manner, e.g., by means of a rotation of the coil head around the *x* axis. Analogous schemes can be devised for more complex cases, where a higher number of unknowns (including for example various mechanical imperfections) of the magnet and of the probe can be found by measuring more flipped configurations.

## 9 Calibration of fixed coils

In this final section we shall discuss the case of fixed coils, where the flux change is determined by the time variation of the external field. Since coil sensitivity factors depend only on the geometry and not on the mechanism of flux change, all results valid for harmonic coils still apply. Considering for simplicity the case of a dipole, integration of the output voltage of a fixed coil while the field is pulsed will give

$$-\frac{1}{w_{eff}} \int_0^t V_c dt = \int_0^L B(s, I(t))ds - \int_0^L B(s, I_0)ds = Bd\ell - Bd\ell_0 . \tag{31}$$

The dynamic nature of the problem entails, however, some important issues related to the reproducibility of the magnetic field in case of cyclical powering. This is affected by a number of factors:

- stability of the power supply, including the accuracy of current measurements;
- hysteresis effects in the iron or in the superconductor, as the case may be. These depend in general upon powering history, temperature or even current ripple (this may shift the magnetic state back and forth along minor hysteresis loops);
- eddy currents and other dynamic effects, which introduce additional time and ramp-rate dependencies (e.g., magnetic aftereffect in the iron [26], decay and snapback in superconducting magnets [27]).

Repeated pre-cycling up to the maximum working currents can help stabilize the magnet onto a reproducible state by way of letting sufficient time for all transient phenomena to die out (a few cycles are often found to be sufficient). Further improvement of stability can be achieved by minimizing other sources of uncertainty, e.g., by starting and stopping the integration in a current range where the power supply is more stable (i.e., far from zero for two-quadrant converters); or choosing an integration time which is an integer multiple of the dominant perturbation, e.g., the 50 Hz from the mains or the ripple of the power supply.

From Eq. (31), we see that a fixed-coil measurement provides only relative flux changes and not absolute values. Nevertheless, such a method is adequate to obtain information on integral multipole field errors, which are by definition a relative quantity (provided that the field distribution at $I = I_0$ is either known or can be safely ignored). We shall now consider the calibration procedure that applies in such a case, taking as an example the curved-coil fluxmeter built for the main dipoles of a hadron therapy synchrotron [28].

The fluxmeter in question, shown in Fig 29, is a flat array of 13 nearly identical and parallel coils lying on the plane where field uniformity has to be measured. Despite the effort expended in winding and sorting, the total surfaces of these coils differ from each other by a few $10^{-3}$, which is larger than the field errors to be measured. This random effect is thought to be mainly due to the deformation of the coils, initially straight and then bent to conform to the curved beam paths through the dipole, and is exacerbated by mechanical handling and temperature changes.

The most efficient way to calibrate out these differences consists in taking one additional coil as a common reference, *provided that* this reference is used to measure exactly the same field as the coil it is being compared to. An elegant way to implement this solution is illustrated in Fig. 30: the reference coil and the array are first placed symmetrically w.r.t. to the mid-plane of the dipole, then the reference is moved sequentially on top of all coils in the array and a measurement in bucked configuration (e.g., series opposition) is taken to obtain

$$\frac{\Phi^j}{\overline{w}_{eff}^j} = \frac{\Phi^{ref}}{\overline{w}_{eff}^{ref}} = -Bd\ell^j + Bd\ell_0^j \quad \Rightarrow \quad \overline{w}_{eff}^j = \overline{w}_{eff}^{ref}\underbrace{\left(1+\frac{\Delta\Phi^j}{\Phi^{ref}}\right)}_{k_j}, \qquad (32)$$

where $\Phi^{ref}$ and $\Phi^j$ are the fluxes measured by the reference and $j$-th array coils respectively during calibration, and $\Delta\Phi^j$ is their difference. When measuring the field uniformity, the relative field differences w.r.t. the central coil (index 0) can therefore be expressed compactly by means of the calibration coefficients $k_j$ as

$$\frac{\Delta Bd\ell^j}{Bd\ell^0} = \frac{\Phi^j}{\Phi^0}\frac{\overline{w}_{eff}^0}{\overline{w}_{eff}^j}-1 = \frac{\Phi^j}{\Phi^0}\frac{k_0}{k_j}-1 \ . \qquad (33)$$

In practical use, during series tests this calibration procedure was found to be necessary several times for each magnet and at least two to three times each month, in order to keep the errors due to field profile differences, temperature drifts, and mechanical instabilities down to an acceptable level.

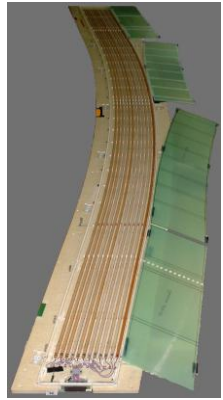

**Fig. 29:** 13-coil array fluxmeter for pulsed-mode measurements of main CNAO synchrotron dipoles

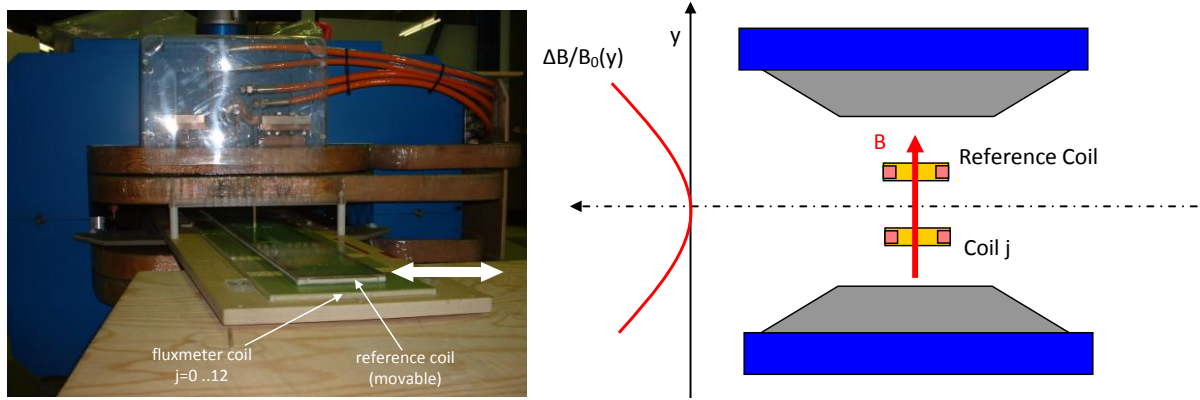

**Fig. 30:** Relative surface calibration procedure for the pulsed-mode fluxmeter. The fluxmeter is inserted in the magnet to be measured alongside an additional, movable reference coil (left). The reference and the calibrated coil are positioned symmetrically in the dipole field (right).

This method can obviously work only if the magnetic field is top–bottom symmetric, i.e., if only the allowed harmonics $B_1$, $B_3$, $B_5$ … are present. Failing that, one could in principle repeat each measurement twice by swapping reference and calibrated coil in the same physical position and then subtracting the results, however, in this way the cycle reproducibility error would have to be taken into account explicitly instead of being automatically cancelled out.

It is important to stress that this calibration procedure cannot be carried out merely by taking one of the coils in the array as the reference, because in this case the measured flux differences would depend upon the unknown differences of field profile, hence the equation system would be underdetermined. Another conceivable alternative consists in the absolute calibration of each coil of the array, in which case the calculation of the desired field errors becomes trivial. Unfortunately, this approach is hardly practical due to the difficulties inherent in the absolute calibration which, as we have seen in Section 7.2, must be carried out in a magnet with the same longitudinal field profile that has been measured with an independent, absolute method. In the author's experience all ordinary options[5] in this respect are either not accurate enough or not cost-effective.

## 10   Conclusions

It is hoped that this lecture has provided a comprehensive (but not exhaustive) overview of the different techniques that can be used to make and calibrate search coils, especially harmonic coils, in order to carry out high-accuracy measurements. It would be tempting to draw up a long final list of recommendations, such as to remember the fundamental importance of a sound mechanical design, or take the utmost care to control the effects of temperature, which is well known as the biggest enemy of precision; however, it is felt that good practice really boils down to common sense and a sound knowledge of physical factors affecting the materials and the electronic equipment. The final message is that, as experimentalists, we should never forget to put beautiful ideas and equations to the test in a variety of environments and conditions, because real-world tests often reserve unexpected surprises.

---

[5] These might include stretched wire techniques (hardly sensitive enough in pulsed mode, and out of the question in the case of a curved magnet); a combination of maps obtained with NMR (magnet centre) and Hall (magnet ends) probes (this would suffer from both the inherent lower accuracy of Hall plates and the error due to the extrapolation from steady-state to dynamic conditions); a high-resolution scan with small coils (probably accurate enough but very time-consuming).

**Acknowledgements**

First and foremost I wish to dedicate this lecture to the late Jacques Billan, who was for three decades the person in charge of CERN's coil manufacturing and calibration laboratory and to whom many of the solutions described in this paper, not to mention much of my own education, are due.

My gratitude goes to present and past team leaders L. Bottura, D. Cornuet, P. Sievers, and L. Walckiers for their guidance and their insight.

I would like to acknowledge gratefully the hard and professional work of present and past magnetic measurement team members, including: R. Beltron Mercadillo, G. Busetta, R. Camus, R. Chritin, D. Cote, G. Deferne, O. Dunkel, J. Dutour, F. Fischer, L. Gaborit, P. Galbraith, J. Garcia Perez, D. Giloteaux, P. Leclere, A. Musso, A. Ozturk, S. Pauletta, S. Sanfilippo, N. Smirnov, S. Stokic, L. Vuffray.

Many thanks also to J. Di Marco (FNAL) and A. Jain (BNL) for valuable discussions and suggestions.